\documentclass[
 reprint,
superscriptaddress,
preprintnumbers,
nofootinbib,
 amsmath,amssymb,
 aps,
]{revtex4-1}

\usepackage{booktabs}
\usepackage{amsmath}
\usepackage{graphicx}
\usepackage{dcolumn}
\usepackage{bm}

\usepackage{booktabs}

\usepackage{slashed}

\usepackage{float}

\usepackage{xcolor}
\usepackage{feynmp-auto}
\definecolor{romared}{RGB}{142,0,28}
\definecolor{tabblue}{RGB}{31, 119, 180}
\definecolor{darkblue}{RGB}{0, 0, 120}
\definecolor{tabred}{RGB}{214, 39, 40}
\definecolor{tabgreen}{RGB}{44, 160, 44}
\definecolor{tabgray}{RGB}{100, 100, 100}
\definecolor{goldenrod}{RGB}{218, 165, 32}
\definecolor{taborange}{RGB}{255, 127, 14}
\definecolor{tabbrown}{RGB}{128, 0, 0}
\definecolor{tabpink}{RGB}{255, 141, 161}
\definecolor{tabpurple}{RGB}{148, 103, 189}

\usepackage[usenames,dvipsnames]{xcolor}
\definecolor{celticsgreen}{rgb}{0, 0.478, 0.2}
\definecolor{celticsgold}{rgb}{0.58823529411, 0.21960784313, 0.1294117647}

\usepackage[colorlinks=true,citecolor=celticsgreen,linkcolor=celticsgreen,urlcolor=celticsgreen,backref=false,pdfborder={0 0 0}]{hyperref}

\begin{document}

\preprint{MIT-CTP/5963}

\title{Kinetic Mixing and the Phantom Illusion:\\ Axion–Dilaton Quintessence in Light of DESI DR2}

\author{Michael W. Toomey}
\email{mtoomey@mit.edu}
\affiliation{Center for Theoretical Physics -- a Leinweber Institute, Massachusetts Institute of Technology, Cambridge, MA 02139, USA}

\author{Ellie Hughes}
\affiliation{Center for Theoretical Physics -- a Leinweber Institute, Massachusetts Institute of Technology, Cambridge, MA 02139, USA}

\author{Mikhail M. Ivanov }
\email{ivanov99@mit.edu}
\affiliation{Center for Theoretical Physics -- a Leinweber Institute, Massachusetts Institute of Technology, Cambridge, MA 02139, USA}
 \affiliation{The NSF AI Institute for Artificial Intelligence and Fundamental Interactions, Cambridge, MA 02139, USA}

\author{James M. Sullivan}
\thanks{Brinson Prize Fellow}
\affiliation{Center for Theoretical Physics -- a Leinweber Institute, Massachusetts Institute of Technology, Cambridge, MA 02139, USA}

\date{\today}

\begin{abstract}
    Recent results from DESI BAO analyses suggest that dark energy may not be a cosmological constant and is in fact dynamical. Furthermore, the data suggest that the equation of state may have been in the phantom regime in the distant past, recently undergoing a phantom crossing. In this work, we investigate whether this preference can be realized within a kinetically mixed axion–dilaton (KMIX) quintessence model, a string-motivated system in which an axion-like field couples exponentially to a dilaton-like (moduli) field. Crucially, KMIX can \emph{appear} phantom in a standard Chevallier–Polarski–Linder (CPL) based analysis while the underlying theory remains non-phantom and stable at all times. To confront the model with data, we develop a fast pipeline based on normalizing flows that (i) learns a theory-informed prior on $(w_0,w_a)$ from KMIX realizations and (ii) provides an inverse mapping from CPL parameters back to the physical KMIX parameters. By importance-sampling pre-computed CPL chains using this framework, we effectively transform generic phenomenological constraints into direct, computationally efficient constraints on the underlying KMIX theory, avoiding the prohibitive cost of full parameter space exploration. We propose that this framework offers a powerful computational shortcut for efficiently constraining a broad swath of theoretical models, transforming generic phenomenological fits into high-fidelity theory-level constraints with negligible computational overhead. Applied to \textit{Planck}+DESI DR2 BAO measurements, our framework finds support for KMIX at $2.5\sigma$ compared to the base CPL fit at $3.1\sigma$, demonstrating that KMIX may account for the DESI preference \emph{without} invoking true phantom behavior. When additionally including Type Ia supernovae data, we find that the preference remains above $3\sigma$ for Union3 and DES~Y5, but drops to $2.1\sigma$ with Pantheon+. The latter, combined with the DESI full-shape power spectrum and bispectrum data, further reduces the preference to $1.7\sigma$. Ultimately, should the DESI deviation persist with future data, KMIX may offer a theoretically well-motivated explanation for the phantom-like signatures inferred from phenomenological fits.
\end{abstract}

\maketitle

\section{Introduction \label{sec:intro}}

Recent results from the Dark Energy Spectroscopic Instrument (DESI) \cite{DESI:2024mwx,DESI:2025zgx} have renewed interest in the possibility that cosmic acceleration may not be driven by a cosmological constant, but by a dynamical dark energy (DE) component. In particular, analyses combining DESI DR2 baryon acoustic oscillation (BAO) data with cosmic microwave background (CMB) measurements have reported a preference for an evolving equation of state in the standard Chevallier--Polarski--Linder (CPL) parameterization \cite{Chevallier:2000qy,Linder:2002et},
\begin{equation}
    w(a) = w_0 + w_a (1-a),
\end{equation}
with a significance $\sim 3\sigma$ \cite{DESI:2025zgx}. Moreover, the inferred time evolution suggests a so-called ``phantom crossing,'' in which the effective equation of state transitions from $w<-1$ at early times to $w>-1$ today \cite{DESI:2025fii}. 
While there is an ongoing debate in the literature
about the significance
of DESI's results,
especially in the context 
of alternative 
DE parameterizations (see e.g.~\cite{Chen:2024vuf,DESI:2024kob,Luongo:2024fww,Shajib:2025tpd,Jiang:2024xnu,Shlivko:2025fgv,Efstathiou:2025tie,Wang:2025vfb, Cheng:2025lod,Wang:2025vfb,Chudaykin:2025aux,deSouza:2025vdv,Toomey:2025xyo,Ong:2025utx}),
the phantom crossing appears
to be substantially preferred
by the data regardless
of adopted parameterizations
and analysis modes~\cite{DESI:2024kob,DESI:2025fii}.

The apparent phantom behavior poses well-known theoretical difficulties. Models that persist in the phantom regime can violate the Null Energy Condition, leading to the possibility for 
ghost instabilities 
\cite{Caldwell:1999ew,Hu:2004kh}.\footnote{See \cite{Caldwell:2025inn} for a recent discussion on the subtleties associated with violation of the Null Energy Condition.} A growing class of explanations which avoid these peculiarities instead posit that the apparent phantom crossing 
is simply 
an artifact of the dark energy 
parameterization
used by DESI.
In particular, it may be avoided in 
non-minimal extensions
of the standard dark matter and/or dark energy sectors.
In these scenarios, interactions between multiple fields, couplings to dark matter, or nontrivial kinetic structures can generate an \emph{effective} phantom signature without introducing ghost-like instabilities \cite{Das:2005yj,Carroll:2004hc,Khoury:2025txd,Huey:2004qv,Brax:2023qyp,Smith:2024ibv,Bedroya:2025fwh,Silva:2025hxw,Wolf:2024eph,Wolf:2024stt,Wolf:2025jed,Goldstein:2025epp,Bottaro:2024pcb,Burgess:2025vxs}. This can be understood in the context of standard analyses for evolving dark energy in the CPL representation where the dark matter content is completely pressureless by construction at all times. 

The limitations of the CPL parameterization in capturing these effects have motivated a return to more physically grounded models of dark energy, in particular those based on scalar-field dynamics. Within the class of \textit{quintessence} models \cite{Peebles:1987ek,Ratra:1987rm}, the dark energy component arises from the slow roll of a scalar field $\phi$ evolving under a potential $V(\phi)$. At the background level, its equation of state is given by
\begin{equation}
    w_\phi = \frac{\phi'^2 - 2a^2 V_\phi}{\phi'^2 + 2a^2 V_\phi},
\end{equation}
which is restricted to $-1\le w_\phi \le 1$ for canonical kinetic terms, precluding genuine phantom behavior. This restriction naturally extends to multi-field systems that are minimally coupled and lack interactions in the kinetic sector.  

If one hopes to evade this bound, one must go beyond canonically normalized scalar fields for quintessence. Several promising frameworks have emerged in which kinetic or potential couplings between scalar fields can mimic phantom-like evolution while preserving theoretical consistency \cite{Smith:2024ibv}. Notably, coupling dark energy to dark matter can modify the effective equation of state through an evolving mass of the dark matter species, allowing $w_{\rm eff} < -1$ even when each field remains well behaved individually. A related and particularly motivated construction arises from the \emph{axion–dilaton} sector of string theory, where moduli and axion fields possess an exponential kinetic coupling inherited from the K\"ahler potential \cite{Bernardo:2022ztc,Alexander:2022own}. This structure naturally yields a kinetically mixed two-field system whose effective dynamics can exhibit apparent phantom behavior \cite{Smith:2024ibv,Toomey:2025mvx}. 

In this work, we investigate this possibility using the \textit{Kinetically Mixed Axion–Dilaton Quintessence} (KMIX) model. This two-field system, in which an axion-like field with a periodic potential interacts with a dilaton-like field via an exponential kinetic coupling, provides a minimal, theoretically motivated mechanism for generating effective phantom signatures. To confront the model with data, we employ a fast inference pipeline based on normalizing flows (NFs)~\cite{Toomey:2024ita}, building on recent applications of simulation-based priors~\cite{Ivanov:2024hgq,Ivanov:2024xgb,Ivanov:2024dgv,Chen:2025jnr}. By learning the bidirectional mapping between the fundamental KMIX parameters and their effective CPL representation, the NF allows us to (i) construct \textit{theory-informed priors} on $(w_0, w_a)$ to efficiently reweight standard phenomenological chains, and (ii) subsequently invert the mapping to derive constraints on the underlying theory space. This strategy facilitates a rigorous comparison between phenomenological fits and physical multi-field scenarios, achieving a dramatic acceleration in inference speed relative to direct sampling of the full model space.

We note that our
NF-based approach to 
theory-informed
importance sampling 
can be applied in a more general context 
to any model whose behavior can be reduced 
to non-linear priors
on the phenomenological
parameters such as energy density or the equation of state coefficients (in this work $w_0$ and $w_a$). This will allow one to easily obtain posteriors
on the model parameters
from the pre-computed
Markov chain Monte Carlo (MCMC) chains, which greatly speeds up 
the analysis of 
physical dark sector
models that can require significantly more computational resources (see, for example, \cite{Toomey:2024ita}).

This paper is organized as follows. In Sec.~\ref{sec:theory}, we introduce the KMIX framework, outlining the string-inspired axion–dilaton dynamics that allow it to generate apparent phantom signatures. Sec.~\ref{sec:priors} presents our end-to-end inference pipeline, detailing both the construction of theory-informed priors on $(w_0, w_a)$ and the inverse mapping used to project observational constraints back into the underlying parameter space. In Sec.~\ref{sec:analysis}, we describe the datasets and statistical methodology used to constrain the model with data, followed by the resulting constraints on the KMIX scenario in Sec.~\ref{sec:results}. Finally, we conclude in Sec.~\ref{sec:conclusion} with a summary of our findings and their broader implications for dark energy model building.

\section{Kinetically Mixed \\Axion–Dilaton System}
\label{sec:theory}

The kinetic coupling between axion- and dilaton-like (moduli) fields arises naturally in string theory compactifications, where moduli and axions share a common K\"ahler potential that governs their field-space metric. In both heterotic and type~II theories, this structure generically leads to exponential factors in the kinetic term of the axion, reflecting its coupling to moduli fields \cite{Bernardo:2022ztc}. After dimensional reduction, one finds a non-canonical field-space metric of the schematic form
\begin{equation}
    \mathcal{L}_{\rm kin} \supset (\partial \chi)^2 + e^{2\chi}(\partial \phi)^2,
    \label{Kahler-structure}
\end{equation}
where $\chi$ and $\phi$ denote the moduli and axion fields, respectively. This exponential dependence---a robust consequence of string compactification and moduli stabilization---implies that kinetic mixing between light scalar fields is a generic feature of the low-energy effective theory \cite{Cicoli:2012sz,Arvanitaki:2009fg,Alexander:2022own,Burgess:2021obw,Smith:2024ibv,Toomey:2025mvx}. For further discussion of the theoretical origin of such couplings, we refer the reader to \cite{Alexander:2022own}.

The phenomenology of such models has recently been explored across several contexts—from addressing the Hubble and large-scale-structure tensions \cite{Alexander:2022own}, to describing ultralight condensate dark matter \cite{Toomey:2025mvx}, and even as a warm inflation mechanism \cite{Berera:2025jbj}. A central feature common to all is the possibility of transferring energy density between the axion and dilaton sectors through the kinetic coupling. In a setting where KMIX plays the role of a two-field quintessence-like model, this transfer can produce an \emph{apparent} phantom equation of state for the moduli field $\chi$ where the effective energy density increases with time. This makes the kinetically mixed axio–dilaton framework a particularly well-suited for explaining the apparent phantom behavior suggested by DESI. 

\begin{figure*}[!t]
    \centering
    \includegraphics[width=0.9\linewidth]{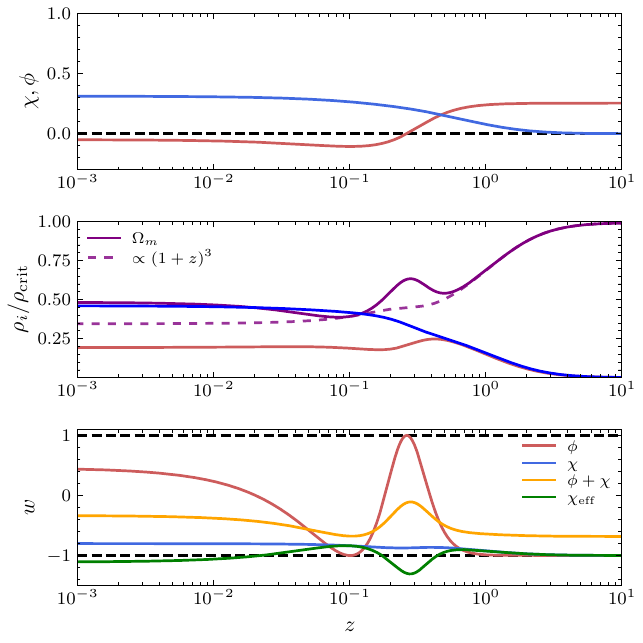}
    \caption{ 
    Example background dynamics for the KMIX model with $\lambda=-1.25$, $\alpha=1.7$, $\log_{10}\chi_i=-4$, $\log_{10}m = -31.8$, $\log_{10}f_\phi = 26.3$, and $\theta_i=3.12$. {\it Top panel:}  Field evolution in units of the reduced Planck mass, with $\phi$ (\textcolor{tabred}{red}) and $\chi$ (\textcolor{tabblue}{blue}) shown relative to the black dashed line marking $\phi, \chi = 0$. Both fields remain frozen at early times and become dynamical once $H \sim m_\phi$. {\it Middle panel:} Evolution of the fractional density for KMIX scalars $\chi$ (\textcolor{tabblue}{blue}) and $\phi$ (\textcolor{tabred}{red}) as well as for all of matter (including the scalar contribution) (solid \textcolor{tabpurple}{purple}). Additionally, we show the evolution of the early-Universe evolved dark matter density (dashed \textcolor{tabpurple}{purple}) which makes clear the impact this model has on the inferred dark matter density at late times.
    {\it Bottom panel:} Evolution of the equation of state for the $\phi$ (\textcolor{tabred}{red}) and $\chi$ (\textcolor{tabblue}{blue}), together with the effective dark-energy equation of state $w_{\chi,{\rm eff}}$ (\textcolor{tabgreen}{green}), as defined in Eq.~\eqref{chi_eff}. The horizontal black dashed lines mark $w=-1$ and $w=1$. For $f(\chi)=e^{\lambda\chi}<1$, consistent with the choice of $\lambda$ used here, the kinetic coupling transfers energy from the axion ($\phi$) to the dilaton ($\chi$), effectively increasing $\rho_\chi$ over time. This energy flow produces an {\it apparent} phantom phase (seen here starting at $z\approx0.5$ where $w_{\chi,{\rm eff}}<-1$) even though the total system satisfies the Null Energy Condition as evidenced from the combined equation of state for $\phi + \chi$ (\textcolor{taborange}{orange}). 
    }
\label{fig:equation-of-state}
\end{figure*}

In this work we specifically have adopted a simplified four-dimensional effective theory inspired by the general structure of Eq.~\eqref{Kahler-structure},
\begin{equation}
    S = \int d^4x\,\sqrt{-g}\,\Big[-\frac{1}{2}(\partial_\mu \chi)^2 - \frac{1}{2}f(\chi)(\partial_\mu \phi)^2 - V(\chi,\phi)\Big],
    \label{eq:action}
\end{equation}
where $\chi$ is a dilaton-like quintessence field and $\phi$ is an axion-like field. The function $f(\chi)=e^{\lambda\chi}$ encodes the exponential kinetic coupling between them, while the potential separates as
\begin{equation}
    V(\chi,\phi)=V_0 e^{-\alpha \chi} + m^2 f_\phi^2 \big[1-\cos(\phi/f_\phi)\big],
\end{equation}
with $(m,f_\phi)$ the axion mass and decay constant, $V_0$ the scale of the $\chi$ potential, and $\alpha$ setting the steepness of the moduli field's potential. This construction adopts the standard potentials for these fields: a periodic cosine for the axion $\phi$, dictated by its discrete shift symmetry, and an exponential for the dilaton $\chi$, characteristic of moduli fields in string effective theories.

This construction captures the essential dynamics of the kinetically mixed axion–dilaton system. The kinetic coupling facilitates energy exchange between the two fields, generating complex late-time evolution—specifically the apparent phantom behavior preferred by DESI—even from simple constituent potentials.\footnote{We emphasize that $\chi$ remains minimally coupled to gravity, ensuring the model naturally evades fifth-force constraints.} We can see this more clearly from the resulting equations of motion which are modified from the standard Klein-Gordon evolution,
\begin{subequations}
\begin{align}
    \ddot{\phi} + \left(3H + \frac{f_\chi}{f} \dot{\chi}\right)\dot{\phi} + \frac{1}{f} V_\phi &= 0, \label{phi_eom}\\
    \ddot{\chi} + 3 H \dot{\chi} - \frac{1}{2}f_\chi \dot{\phi}^2 + V_\chi &= 0.
    \label{chi_eom}
\end{align}
\end{subequations}
Inspecting these relations, we see that the axion's equation of motion acquires an effective friction term proportional to $\lambda \dot\chi$, along with a modulation of its potential by $1/f$. Conversely, the dilaton equation of motion includes a source or sink term (dependent on the sign of $\lambda$) driven by the kinetic energy of the axion, $\frac{1}{2}\lambda f \dot\phi^2$. Beyond the background dynamics, KMIX also modifies the evolution of linear perturbations. For a detailed treatment, we refer the reader to Ref.~\cite{Alexander:2019rsc}.

Its also instructive to write down the modification to the energy densities and pressures in this model for both fields. For $\chi$ they are,
\begin{equation}
        \rho_\chi = \frac{1}{2}\dot\chi^2 + V(\chi),
\end{equation}
\begin{equation}
        p_\chi = \frac{1}{2}\dot\chi^2 - V(\chi),
\end{equation}
and for $\phi$,
\begin{equation}
        \rho_\phi = \frac{1}{2}f\dot\phi^2 + V(\phi),
\end{equation}
\begin{equation}
        p_\phi = \frac{1}{2}f \dot\phi^2 - V(\phi).
\end{equation}
It is easy to show from here that this model will not violate the NEC, that is, $-1\leq w_{\phi + \chi} \equiv p_{\phi + \chi} /\rho_{\phi + \chi}  \leq 1$. However, if we assume
dark matter to be purely pressureless ($w=0$), 
which is done in most DESI-based analyses, 
one can have an apparent phantom equation of state in the KMIX model. By absorbing the kinetic coupling terms into the dynamics of the $\chi$ field, we define an effective equation of state for the dark energy sector:
\begin{equation}\label{chi_eff}
    w_{\chi,\rm eff} = \frac{\dot\chi^2 - 2V(\chi) + (f(\chi) - 1)\dot\phi^2}{\dot\chi^2 + 2V(\chi) + (f(\chi) - 1)\dot\phi^2}.
\end{equation}
Inspection of Eq.~\eqref{chi_eff} reveals the mechanism for phantom mimicry: whenever the coupling function satisfies $f(\chi) < 1$, the kinetic contribution in the numerator is suppressed relative to the denominator, making the phantom regime $w_{\chi,\rm eff} < -1$ dynamically accessible.

We illustrate this behavior in Fig.~\ref{fig:equation-of-state} for a benchmark model selected to resemble the best-fit DESI expansion history. Here, the effective phantom crossing is triggered as the axion exits the slow-roll regime and is released from Hubble friction; consequently, we anticipate that reproducing this feature requires an axion mass $m \gtrsim H_0 \sim 10^{-33}$~eV. Crucially, the figure also confirms that the total system satisfies the Null Energy Condition at all times: the combined equation of state evolves from $w_{\rm total} \approx -1$ during the initial slow-roll phase toward larger (but still negative) values today, demonstrating that the phantom signature is strictly an effective phenomenon arising from the two-field interaction.

\section{Generating Efficient constraints on the KMIX Model} \label{sec:priors}

In this section, we outline the technical framework used to obtain efficient constraints on the KMIX model. Our approach employs a two-stage importance--sampling pipeline that transforms the existing CPL-based MCMC chains into samples consistent with the KMIX theory space. First, we impose KMIX-informed priors on the cosmological parameters using a conditional normalizing flow trained on theory-space data. Second, we use a separate normalizing flow to learn the non-linear mapping between the CPL $(w_0, w_a)$ representation and the underlying KMIX parameter space. Our NFs, a class of generative machine-learning models that learn bijective transformations between complex target distributions and simple base distributions \cite{2019arXiv191202762P}, allow us to reweight the CPL chains into high-fidelity KMIX chains without requiring a full MCMC run for the KMIX model.

\subsection{Theory-Informed Priors for KMIX}
\label{sec:kmix_priors_text}

\begin{table}[!t]
\centering
\caption{Prior ranges adopted for the KMIX model parameters and used to generate effective prior on ($w_0,w_a$) -- see Fig.~\ref{fig:placeholder}.}
\label{tab:kmix_priors}
\begin{tabular}{lc}
\hline\hline
\textbf{Parameter} & \textbf{Prior Range} \\
\hline
$\lambda$                & $\mathcal{U}(-3,\,0)$ \\
$\log_{10}(\chi_i)$      & $\mathcal{U}(-6,\,1)$ \\
$\alpha$                 & $\mathcal{U}(0,\,10)$ \\
$\theta_i$               & $\mathcal{U}(0,\,\pi)$ \\
$\log_{10}(f_\phi/{\rm eV})$ & $\mathcal{U}(25,\,28)$ \\
$\log_{10}(m/{\rm eV})$  & $\mathcal{U}(-33,\,-30)$ \\
$H_0~[{\rm km\,s^{-1}\,Mpc^{-1}}]$ & $\mathcal{U}(50,\,90)$ \\
$\omega_{\rm b}$          & $\mathcal{U}(0.0210,\,0.0245)$ \\
$\omega_{\rm cdm}$        & $\mathcal{U}(0.100,\,0.140)$ \\
\hline\hline
\end{tabular}
\end{table}

\begin{figure*}
    \centering
    \includegraphics[width=0.9\linewidth]{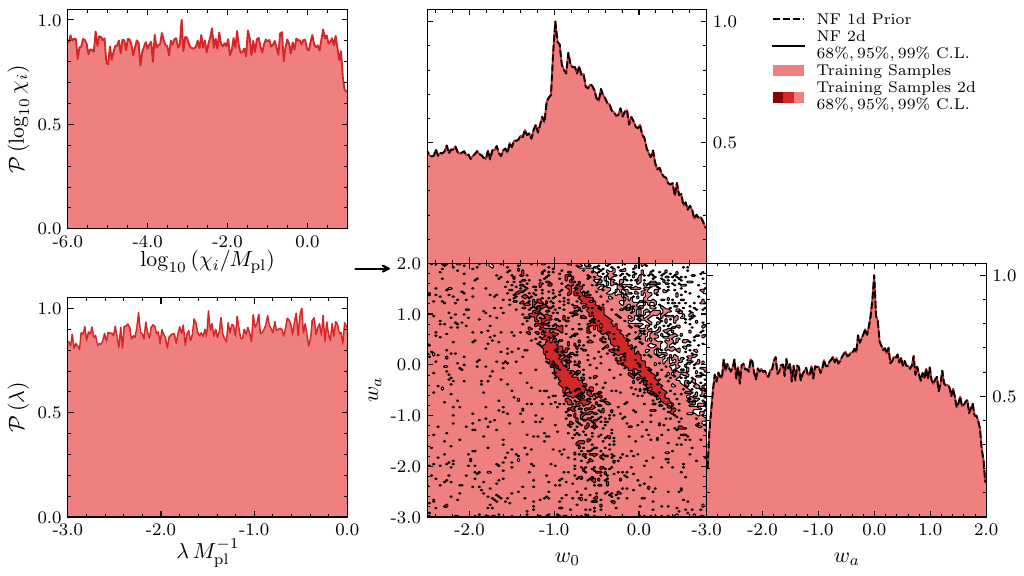}
    \caption{Theory-informed prior distributions for the KMIX model, derived through normalizing flow (NF) training using samples from the parameter ranges listed in Table~\ref{tab:kmix_priors}. {\it Left panel:} Raw training samples drawn from broad flat priors on the KMIX theory parameters, shown here for representative quantities $\chi_i$, and $\lambda$, demonstrating full coverage of the prior volume. {\it Right panel:} Triangle plot of the corresponding mapped $(w_0, w_a)$ distributions obtained through the CPL-matching procedure. Light red points show the raw training data, while increasingly darker contours indicate the 99\%, 95\%, and 68\% confidence regions of the resulting NF prior. Black curves denote the NF-learned priors: dashed lines for the 1D marginalized distributions and solid contours for the 2D joint distribution.}
    \label{fig:placeholder}
\end{figure*}

Since our base chains were sampled under uniform priors on ($w_0,w_a$), the CPL posterior does not directly reflect the probability density implied by the KMIX model. However, by reweighting these chains using a KMIX-induced prior in $(w_0, w_a)$ one can alternatively provide a computationally efficient surrogate for the posterior that would be obtained from a full MCMC performed directly in KMIX theory space. In this sense, importance resampling the CPL chains with a KMIX prior for ($w_0,w_a$) functions as the theoretical lens through which one can interpret the viability of KMIX. There are of course limitations to this approach, but the trade off is efficient generation of constraints without a need for computationally expensive MCMC runs, which can become intractable for models with detailed UV physics (see discussions in \cite{Toomey:2024ita} in the context of early dark energy).

To generate the KMIX informed prior on ($w_0,w_a$) we follow the methods developed in \cite{Toomey:2024ita,Toomey:2025xyo} to generate theory informed priors on phenomenological parameters with NFs. On the theory side, our prior choices for the KMIX model ($\lambda$, $f_\phi$, etc.) are guided by string-theoretic consistency and landscape statistics while remaining deliberately broad to capture all phenomenologically viable regions. On the axion side, quantum-gravity arguments against exact global symmetries~\cite{Kallosh:1995hi,Banks:2003sx}, the axion weak gravity conjecture (WGC)~\cite{Arkani-Hamed:2006emk,Rudelius:2015xta,Brown:2015iha,Hebecker:2015zss}, and explicit compactifications~\cite{Svrcek:2006yi,Conlon:2006tq,Cicoli:2012sz} bound the decay constant to be sub-Planckian and imply an instanton expansion with action $S \sim c M_{\rm pl}/f_\phi$ and $c=\mathcal{O}(1)$, leading to the familiar harmonic structure in $V(\theta)$. Convergence and dominance of the leading mode require $M_{\rm pl}^4 e^{-c M_{\rm pl}/f_\phi}\ll m^2 f_\phi^2$, equivalent to $f_\phi \log\!\left(M_{\rm pl}^4/f_\phi^2m^2\right)\lesssim c M_{\rm pl}$, consistent with refined WGC bounds~\cite{Rudelius:2022gyu}. The Swampland Distance Conjecture (SDC) likewise disfavors super-Planckian excursions~\cite{Ooguri:2006in,Baume:2016psm,Klaewer:2016kiy,Blumenhagen:2017cxt,Scalisi:2018eaz}. Beyond these consistency limits, statistical studies of flux compactifications indicate that both the axion mass and decay constant are approximately log-uniform over many decades~\cite{Broeckel:2021dpz,Halverson:2019cmy,Mehta:2020kwu,Mehta:2021pwf,Demirtas:2021gsq}.
Informed by these theoretical constraints, we take $\lambda$, $\log_{10}\chi_i$, $\alpha$, and $\theta_i$ to be uniform over the ranges shown in Table~\ref{tab:kmix_priors}, and we impose log-uniform priors on the axion mass and decay constant by sampling $\log_{10}(m/{\rm eV})$ and $\log_{10}(f/{\rm eV})$ uniformly over the quoted intervals.

However, to construct theory-informed priors on ($w_0,w_a$) for the KMIX model, we need to know the mapping between this prior in theory space into $w_0,w_a$. To do so, we follow the procedure outlined in~\cite{Toomey:2024ita} and \cite{Toomey:2025xyo}, adapted here for the kinetically mixed axion–dilaton system implemented in our own modified implementation of \texttt{CLASS}~\cite{Lesgourgues:2011re}. The first step requires generating a data set of parameters $\theta_{\rm CPL}$ for each set of theory parameters $\theta_{\rm KMIX}$. For each draw from the theory parameter space defined in Tab.~\ref{tab:kmix_priors}, we solve for the expansion history $H_{\rm KMIX}(z)$ and compare it to the CPL prediction by computing the maximum fractional deviation,
\begin{equation}
    E_H = \underset{z<4}{\max}\left|
    \frac{H_{\rm CPL}(z) - H_{\rm KMIX}(z)}{H_{\rm KMIX}(z)}\right|,
    \label{eq:EH_kmix}
\end{equation}
where the range $z<4$ is chosen to be inclusive of all redshifts constrained by DESI~\cite{DESI:2024uvr,DESI:2024aqx}. However, because the phenomenological CPL parametrization cannot perfectly reproduce the complex dynamics of the physical KMIX model, optimizing Eq.~\eqref{eq:EH_kmix} yields a degeneracy of approximate solutions rather than a single, perfect mapping -- see related discussions in \cite{Toomey:2025xyo}.

To address this inherent ambiguity and construct meaningful prior distributions, we adopt a probabilistic approach that naturally propagates the fitting uncertainty into our final priors (see Ref.~\cite{Toomey:2025xyo} for further details). Rather than selecting a single best-fit point, we generate a probability distribution over the $(w_{0},w_{a})$ plane for each model realization, weighted by the inverse square of the matching error, $P(w_{0},w_{a})\propto1/E_{H}^{2}$. This inverse-variance weighting ensures that regions of parameter space providing poor fits to the quintessence dynamics ($E_H \gg 0$) are assigned lower probability, while the spread of the distribution captures the degeneracy and uncertainty inherent in mapping the physical theory to the effective CPL representation.

Repeating this procedure over a large ensemble of parameter draws generates a training set of $(w_0, w_a, \omega_{\rm b}, \omega_{\rm c}, H_0)$ samples, from which we learn the conditional density
\begin{equation}
    P_{\rm th}(w_0,w_a \mid \omega_{\rm b},\omega_{\rm c},H_0)
\end{equation}
using a conditional normalizing flow. In this work, we simulate $N \simeq 100{,}000$ realizations by sampling the parameter ranges outlined in Table~\ref{tab:kmix_priors}.  Once trained, the NF can be used as a smooth, theory-informed prior that can be directly applied to reweight existing MCMC chains in $w_0w_a$CDM, as described in the previous section. For further implementation details, see \cite{Toomey:2024ita}. 

Figure~\ref{fig:placeholder} illustrates the learned theory-informed priors for $(w_0, w_a)$, conditioned on the best-fit \textit{Planck} 2018 $\Lambda$CDM cosmology, alongside the raw training samples generated from KMIX realizations. The close agreement demonstrates the NF's ability to accurately capture the complex, non-Gaussian structure of the mapping from theory space. Markedly distinct from single-field quintessence models, which result in priors strongly peaked near $(w_0, w_a) \simeq (-1,0)$~\cite{Toomey:2025xyo}, the KMIX framework yields a significantly broader distribution. This resulting prior volume is qualitatively similar to the standard uniform priors employed in CPL analyses~\cite{DESI:2024mwx}, particularly along the $w_a$ direction. Consequently, we anticipate that the importance weights $W_i$ (defined in Eq.~\ref{IRW}) will typically remain close to unity, leading to conservative shifts in the final posterior constraints.

\subsection{Inverse Mapping to KMIX Theory Parameters}
\label{subsec:inverse_mapping}

While the normalizing flow constructed above provides a theory-informed prior on $(w_0, w_a)$ for use in data analyses, it is often more informative to express cosmological constraints directly in terms of the underlying parameters of the KMIX model itself. The CPL parameters, though useful as a phenomenological summary for comparison, compress a high-dimensional theoretical space into a two-parameter effective description that obscures physical correlations between model parameters. To recover these relations, we train a second normalizing flow that learns the inverse mapping from the best-fit CPL parameters $(w_0, w_a)$ back to the fundamental KMIX inputs. Specifically, we train a conditional normalizing flow to approximate the probability distribution
\begin{equation}
    P_{\mathrm{inv}}(\lambda,\,\chi_i,\,\alpha,\,\theta_i,\,f_\phi,\,m \mid w_0,\,w_a, H_0, \omega_{\rm b}, \omega_{\rm cdm}).
\end{equation}
Once trained, this flow enables us to take the reweighted chains described above and infer the underlying KMIX model parameters $\theta_{\rm KMIX}$, providing a direct bridge between the baseline phenomenological analysis and the underlying high-energy theory.

This approach offers three key advantages. First, it connects cosmological data directly to physically meaningful parameters, facilitating direct comparison with theoretical expectations such as Swampland or Weak Gravity Conjecture bounds. Second, it recovers parameter correlations and degeneracies that are otherwise obscured by the $(w_0, w_a)$ compression, providing a clearer physical interpretation of the inferred dark energy dynamics. Finally, this method yields a massive increase in inference speed: whereas direct MCMC sampling of complex particle models can require years of aggregate CPU compute time to converge, our pipeline requires only $\mathcal{O}(1)$ hour to generate the training dataset with \texttt{CLASS} and mere minutes to train the normalizing flow, enabling rapid high-dimensional parameter reconstruction without rerunning the full analysis.

\begin{figure}[!t]
    \centering
    \includegraphics[width=\linewidth]{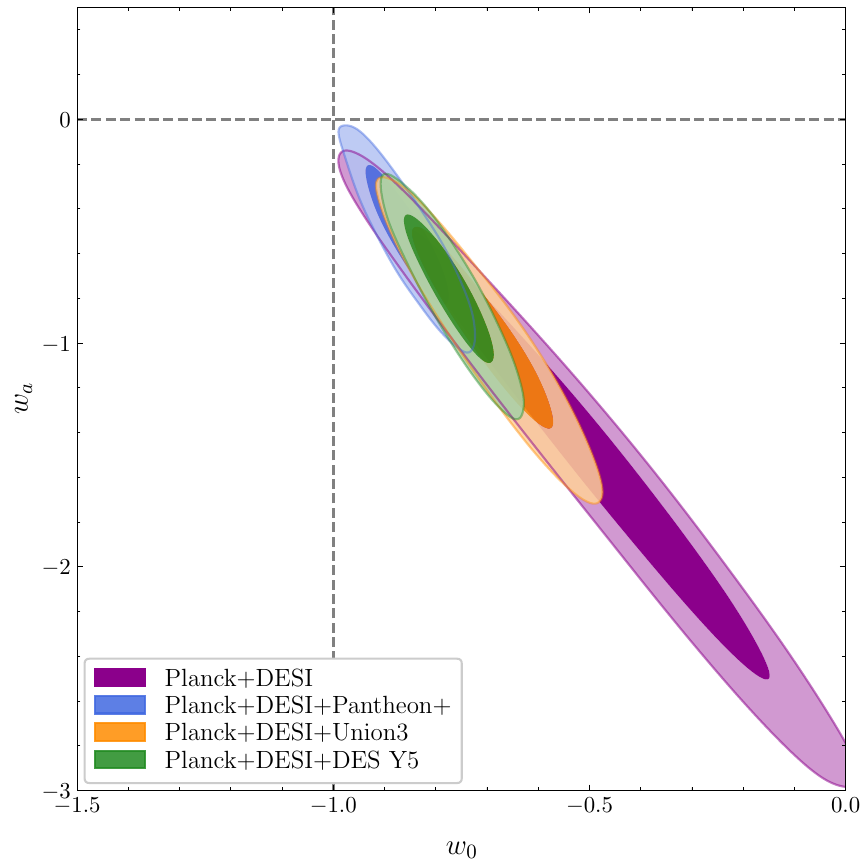}
    \caption{Two-dimensional posterior distributions in the $(w_0,\,w_a)$-plane at the 68\% and 95\% confidence levels for the kinetically mixed axion--dilaton (KMIX) model. The analysis combines \textit{Planck} PR4 CMB anisotropies (with lensing) and DESI DR2 BAO, shown in \textcolor{tabpurple}{purple}, and separately includes Type~Ia supernova datasets from Pantheon+ (\textcolor{tabblue}{blue}), Union~3 (\textcolor{taborange}{orange}), and DES~Y5 (\textcolor{tabgreen}{green}). The gray dashed lines denote the $\Lambda$CDM limit, $(w_0,\,w_a)=(-1,\,0)$.
}
    \label{fig:cmb-bao-kmix}
\end{figure}

\begin{figure*}
    \centering
    \includegraphics[width=\linewidth]{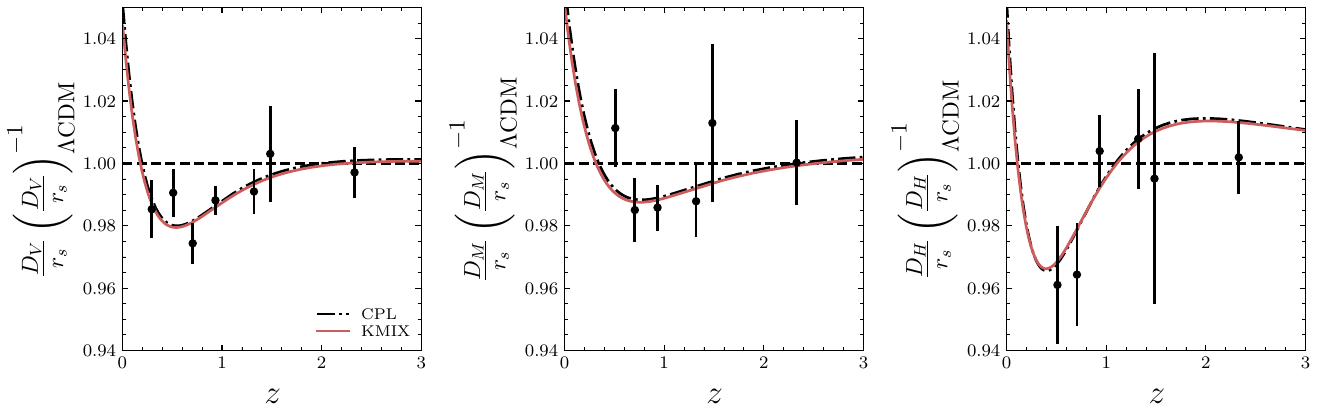}
    \caption{Comparison of best-fit predictions for DESI DR2 BAO distance scales relative to the best-fit {\it Planck} $\Lambda$CDM baseline. The panels show the volume-averaged distance $D_V/r_d$ (left), the transverse comoving distance $D_M/r_d$ (center), and the Hubble distance $D_H/r_d$ (right). Black points indicate DESI DR2 measurements with $1\sigma$ errors. The \textcolor{tabred}{red} line corresponds to the best-fit KMIX model, plotted using the corresponding $(w_{0}^{\rm (eff)}, w_{\rm a}^{\rm (eff)})$, while the black dot-dashed line represents the best-fit standard CPL value.}
    \label{fig:bao-compare}
\end{figure*}

\section{Analysis and Data Sets}\label{sec:analysis}

For our base ($w_0,w_a$) chains, which we use to construct our KMIX constraints as described in the previous section,  we analyze three primary dataset configurations: (1) \textit{Planck} CMB data combined with DESI DR2 BAO measurements, 
(2) CMB+BAO further supplemented with Type Ia supernovae (SNe) likelihoods,
and (3) CMB+BAO+SNe (from Pantheon+) 
enhanced with the full-shape (FS)
galaxy power spectrum
and bispectrum likelihood
from DESI DR1~\cite{Chudaykin:2025aux}. 
For the CMB constraints, we use the \textit{Planck} PR4 data release, incorporating NPIPE-based high- and low-$\ell$ temperature and polarization likelihoods together with the PR4 lensing likelihood~\cite{Efstathiou:2019mdh,Rosenberg:2022sdy,Carron:2022eyg}. This release provides the most internally consistent and robust \textit{Planck} measurements to date, offering improved foreground mitigation and noise modeling relative to the legacy release.
BAO constraints are taken from the latest DESI DR2 measurements \cite{DESI:2025zgx}, which deliver the most precise distance-scale determinations across a wide redshift range. For Type Ia supernovae, we include three independent compilations—Pantheon+~\cite{Scolnic:2021amr,Brout:2022vxf}, Union3~\cite{Rubin:2023jdq}, and DES Year~5 (DES Y5)~\cite{DES:2024jxu}—to test the robustness of our results across distinct samples. 
We note that a recent recalibration of DES~Y5 has reduced the evidence for dynamical dark energy~\cite{Popovic2025:Dovekie_1,Popovic2025:Dovekie_2}.

All likelihoods except FS are implemented within the \texttt{Cobaya} framework~\citep{Torrado:2020dgo}, using its default configurations for supernova analyses. 
Cosmological observables are modeled using the default CPL implementation in the Einstein–Boltzmann solver \texttt{CLASS}~\cite{Blas:2011rf}, including perturbations for likelihood evaluation. Following~\cite{DESI:2025zgx}, we adopt broad uniform priors on the parameter set $(\omega_{\rm c},\omega_{\rm b}, 100\theta_{\rm MC}, \log(10^{10}A_{\rm s}), n_{\rm s}, \tau, w_0, w_a)$—comprising the six standard $\Lambda$CDM parameters plus the two CPL parameters—with $w_0 \in \mathcal{U}[-3,1]$ and $w_a \in \mathcal{U}[-3,2]$. Posterior distributions are sampled via MCMC using the Metropolis–Hastings algorithm~\cite{2005math......2099N}, and chains are run until achieving Gelman–Rubin convergence~\cite{10.1214/ss/1177011136} with $R - 1 < 0.01$. We report one-dimensional marginalized posteriors as means with 68\% minimum credible intervals, analyzing all MCMC outputs using \texttt{GetDist}~\cite{Lewis:2019xzd}. 

As far as the full-shape
galaxy and quasar power spectrum and bispectrum data are concerned, 
we use the MCMC chains 
obtained with the custom 
DESI FS likelihood
of~\cite{Chudaykin:2025aux}
obtained using methodologies 
from~\cite{Ivanov:2019pdj,Chudaykin:2020aoj,Chudaykin:2020ghx,Ivanov:2021kcd,Ivanov:2021zmi,Philcox:2021kcw,Chudaykin:2022nru} (see also~\cite{DAmico:2019fhj,DAmico:2020kxu,Chen:2021wdi,Zhang:2021yna,Chudaykin:2024wlw}) and
supplemented 
with simulation-based priors~\cite{Ivanov:2024jtl,Ivanov:2025qie,Chen:2025jnr} (see also~\cite{Sullivan:2021sof,Akitsu:2024lyt,Cabass:2024wob,Zhang:2024thl,DESI:2025wzd}). Details of this analysis 
can be found in~\cite{Chudaykin:2025auz}.

To obtain constraints on the KMIX model without running computationally expensive chains directly in the high-dimensional theory space, we employ importance sampling on the CPL chains. The target posterior for the KMIX model, $\mathcal{P}_{\rm KMIX}$, is related to the distribution sampled by our base MCMC, $\mathcal{P}_{\rm base}$, via the ratio of their priors. Specifically, for a sample point $i$ with parameters $\mathbf{\theta}_i = \{w_{0,i}, w_{a,i}, \mathbf{\theta}_{{\rm bg},i}\}$, the importance weight $W_i$ is given by:
\begin{equation}
    W_i = \frac{P_{\rm th}(w_{0,i}, w_{a,i} \mid \mathbf{\theta}_{{\rm bg},i})}{\pi_{\rm uni}(w_{0,i}, w_{a,i})},
    \label{IRW}
\end{equation}
where $\pi_{\rm uni}$ is the flat prior used in the base CPL analysis, and $P_{\rm th}$ is the theory-informed prior learned by the conditional NF described in the last section. The background parameters $\mathbf{\theta}_{\rm bg}$ (e.g., $H_0$, $\omega_m$) are included in the conditioning to ensure that correlations inherent to the specific dataset (e.g., the geometric degeneracy in BAO) are correctly propagated into the KMIX constraints. Lastly, we also use the inverse NF described in the last section to add the KMIX model parameters $\theta_{\rm KMIX}$ as derived parameters to generate our high fidelity KMIX chains.

\section{Results \label{sec:results}}

\begin{figure}[!t]
    \centering
    \includegraphics[width=\linewidth]{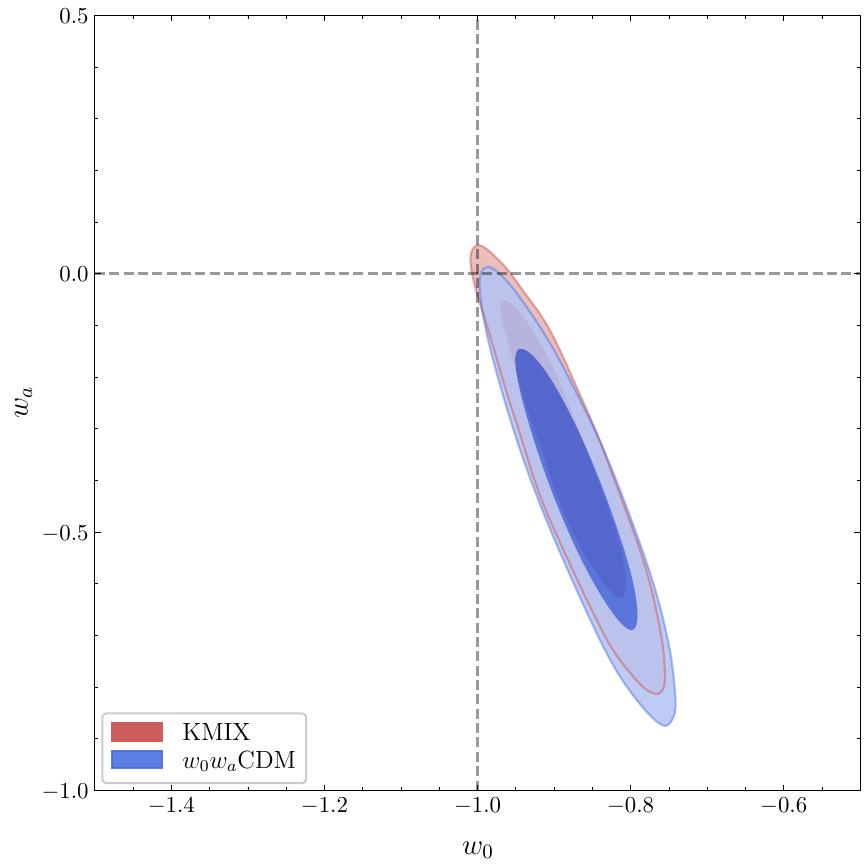}
    \caption{Two-dimensional posterior distributions in the {\it projected} $(w_0,\,w_a)$-plane at the 68\% and 95\% confidence levels from the analysis~\cite{Chudaykin:2025aux,Chudaykin:2025auz}
    based on DESI DR2 BAO, \textit{Planck} CMB, Pantheon+ SNe, and DESI DR1 full-shape data. Results are shown for the KMIX (\textcolor{tabred}{red}) and $w_0w_a$CDM (\textcolor{tabblue}{blue}) models. The gray dashed lines indicate the $\Lambda$CDM limit, $(w_0,\,w_a)=(-1,\,0)$. While both models suggest consistency (with $\Lambda$CDM within $\sim\!2\sigma$) KMIX exhibits a slightly weaker apparent deviation.
}
    \label{fig:fs_constraint}
\end{figure}

\subsection{Baseline DESI Analysis}

We begin with the baseline combination of \textit{Planck}~PR4 CMB anisotropies and DESI~DR2 BAO distance-scale measurements. The resulting constraints for KMIX, mapped into the $(w_0,w_a)$ plane, are shown in Fig.~\ref{fig:cmb-bao-kmix}. These contours represent the range of effective CPL parameters that arise from KMIX cosmologies capable of matching the observed expansion history.
For this dataset combination, KMIX exhibits a moderate preference for evolving dark energy relative to the $\Lambda$CDM limit. As summarized in Table~\ref{tab:sigma_values}, the effective deviation corresponds to $2.5\sigma$, indicating that the two-field axion--dilaton system readily produces the mild phantom-crossing behavior favored by DESI. Importantly, the KMIX contours occupy the same qualitative region as the CPL fit—shifted away from $(w_0,w_a)=(-1,0)$ and lying in the quadrant associated with $w(a)<-1$ at intermediate redshifts—demonstrating that the model naturally reproduces the broad evolution pattern inferred from the data. This differs from the results for canonical scalar fields where there is minimal support for phantom dynamics (e.g., see Fig.~3 in \cite{Toomey:2025xyo}). 

When Type~Ia supernova samples are incorporated, the KMIX preference strengthens for two of three SNe samples, in line with expectations from the CPL analyses. Adding Pantheon+ weakens the preference to $2.1\sigma$ while Union3 and DES~Y5 shifts contours further from $\Lambda$CDM, with nominal deviations at $3.1\sigma$ to $3.4\sigma$, respectively. These results track the same hierarchy seen in CPL fits: high-redshift supernovae (Union3 and DES~Y5) reinforce the DESI pull toward evolving dark energy more strongly than Pantheon+. In each case, the KMIX contours remain consistent with the CPL-derived region but tend to lie slightly closer to the $\Lambda$CDM boundary, reflecting the fact that KMIX realizes a more restricted subset of phantom-like trajectories. This can also be understood from the differences in the effective prior on ($w_0,w_a$), i.e. Fig.~\ref{fig:placeholder}, relative to the standard CPL analysis with uniform $(w_0,w_a)$ priors.

In Fig.~\ref{fig:placeholder_2} of the Appendix we see that the KMIX posteriors select a region of parameter space that aligns well with what one would reasonably expect from a kinetically mixed axion–dilaton model that exhbitis apparent phantom behavior. The axion is pushed toward an initial angle close to $\pi$, placing it near the top of its potential and naturally giving rise to stable slow-roll dynamics. The dilaton begins well below the Planck scale, consistent with the expectations of the Distance Conjecture and with staying in the regime where the exponential kinetic coupling remains trustworthy. The mixing parameter $\lambda$ comes out to be order unity—precisely the range suggested by explicit constructions—and the axion mass sits modestly above $H_0$, as one would anticipate for a field whose late-time evolution drives the effective phantom crossing. Meanwhile, the decay constant remains comfortably sub-Planckian, with $\log_{10}(f_\phi/{\rm eV})\simeq 25.9$, avoiding the usual quantum-gravity tensions associated with $f_\phi \sim M_{\rm pl}$. Altogether, the preferred KMIX parameters form a coherent and theoretically sensible picture, and the resulting best-fit expansion history provides an excellent match to the DESI-preferred CPL behavior. This is illustrated in Fig.~\ref{fig:bao-compare}, where the KMIX prediction for the BAO closely tracks the best-fit $w_0,w_a$CDM from the DESI analysis. 

\begin{table}[!t]
\centering
\caption{Significance of deviations from $\Lambda$CDM in the KMIX model for different datasets and in the CPL model. }
\vspace{4pt}
\label{tab:sigma_values}
\begin{minipage}{\columnwidth}
\centering
\normalsize
\setlength{\tabcolsep}{8pt}
\begin{tabular}{lcc}
\hline\hline
\textbf{Dataset} & \textbf{KMIX} & \textbf{CPL} \\
\hline
Planck+DESI              & $2.5\sigma$ & $3.1\sigma$ \\
Planck+DESI+Pantheon+    & $2.1\sigma$ & $2.5\sigma$ \\
Planck+DESI+Union3       & $3.1\sigma$ & $3.9\sigma$ \\
Planck+DESI+DES~Y5       & $3.4\sigma$ & $4.0\sigma$ \\
\hline
Full Shape               & $1.7\sigma$ & $2.0\sigma$ \\
\hline\hline
\end{tabular}
\end{minipage}
\end{table}

\subsection{Full-shape constraints}

We now extend our analysis to include the DESI full-shape galaxy power spectrum and bispectrum, combined with the baseline BAO, \textit{Planck}~PR4 CMB, and Pantheon+ supernova constraints. As established in standard CPL studies, the FS data exerts a strong pull toward the $\Lambda$CDM limit, reducing the preference for evolving dark energy from $2.5\sigma$ to $2.0\sigma$ relative to geometric fits~\cite{Chudaykin:2025aux}.

Figure~\ref{fig:fs_constraint} demonstrates that this data-driven convergence is even more pronounced for the KMIX model. Starting from a preference of $2.1\sigma$ for the corresponding geometric combination (with Pantheon+), the inclusion of FS data further reduces the effective deviation to $1.7\sigma$ (see Table~\ref{tab:sigma_values}). While the FS data is the primary driver shifting both posteriors toward $(w_0, w_a)=(-1,0)$, the KMIX contours retract further because the standard CPL parameterization possesses significantly more prior volume in the deep phantom regime to support the non-$\Lambda$CDM limit of the parameter space.

Figure~\ref{fig:placeholder_3} in the Appendix presents the constraints on the underlying KMIX parameters derived from the full-shape analysis. While the preferred regions of parameter space remain qualitatively consistent with the baseline geometric results, the marginalized posteriors for the physical parameters notably broaden. This degradation in constraining power is a direct consequence of the data shifting the posterior closer to the $\Lambda$CDM limit: as the preferred phenomenological solution approaches $w(z) \approx -1$, the leverage to distinguish specific dynamical signatures of the kinetic coupling diminishes. Consequently, we observe increased posterior support for axion masses near $m \sim 10^{-33}$~eV, corresponding to a regime where the field remains frozen in slow roll, thereby suppressing the unique phantom-like dynamics that would otherwise distinguish the model from a cosmological constant.

\begin{figure}
    \centering
    \includegraphics[width=\linewidth]{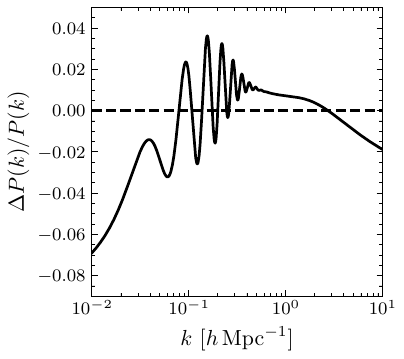}
    \caption{Fractional difference in the linear matter power spectrum between $w_0w_a$CDM and the KMIX power spectrum using parameters calculated from the inverse mapping where $\Delta P = P_{KMIX} - P_{CPL}$.}
    \label{fig:placeholder_4}
\end{figure}

While the CPL parametrization can effectively mimic the background expansion history of the KMIX model, the degeneracy is broken at the level of perturbations. Figure~\ref{fig:placeholder_4} illustrates the fractional difference in the linear matter power spectrum, $P(k)$, between the full KMIX theory and a $w_0w_a$CDM model with parameters derived from our inverse mapping. The visible deviations arise because the scalar fields in KMIX possess internal dynamics and clustering properties (e.g., $c_s^2 \neq 1$) that are not captured by the standard fluid approximation of CPL. 
In particular, the KMIX model produces an additional $\sim 5\%$ suppression of power on large scales ($k\lesssim 0.1~h\text{Mpc}^{-1}$), consistent with the previous BOSS full-shape 
measurements~\cite{Philcox:2021kcw,Chen:2024vuf},
and $\approx 1\%$ enhancement of power on
 small scales 
$k\approx  1~h\text{Mpc}^{-1}$,
which can be probed with the Lyman-$\alpha$
forest data~\cite{Goldstein:2023gnw,Ivanov:2024jtl}.
This demonstrates that geometric probes like BAO are insufficient to fully distinguish the models on their own. Instead, high-precision measurements of the growth of structure are required. 
Consequently, future high-volume spectroscopic surveys such as Spec-S5/MegaMapper~\cite{Spec-S5:2025uom} will be essential for identifying these specific signatures and distinguishing physical multi-field quintessence models like KMIX from phenomenological parametrizations.

\section{Conclusion} \label{sec:conclusion}

In this work, we have examined whether the apparent phantom-like behavior suggested by DESI BAO measurements can be understood within a concrete, theoretically motivated framework, in contrast to the phenomenological CPL parameterization. To this end, we studied the cosmology of a two-field system inspired by the most general structure of the axion--dilaton sector of string axions. In this scenario, an axion-like field is kinetically coupled to a dilaton-like field. The fields play the role of a two-field quintessence model, which further enables an energy transfer between the two sectors. This mechanism naturally generates an \emph{effective} phantom phase in the dark-energy equation of state, even though the total system always remains non-phantom and satisfies the Null Energy Condition.

Mapping KMIX into the $(w_0,w_a)$ plane allows for a direct comparison with DESI analyses of the background expansion history. For the baseline \textit{Planck}~PR4 + DESI~DR2 BAO dataset, we find an effective $2.5\sigma$ preference for KMIX relative to $\Lambda$CDM. 
The corresponding KMIX contours, shown in Fig.~\ref{fig:cmb-bao-kmix}, lie in the same quadrant favored by CPL analyses—shifted away from $(w_0,w_a)=(-1,0)$, toward the region associated with an apparent phantom crossing—illustrating that KMIX naturally reproduces the qualitative evolution preferred by the data.  When Type~Ia supernovae are included, this preference strengthens to the $2.1$--$3.4\sigma$ level depending on the compilation (Pantheon+, Union3, or DES~Y5), with higher-redshift samples driving the largest deviations. In each case, the KMIX contours track the CPL posteriors but tend to sit closer to the $\Lambda$CDM boundary, consistent with the fact that only a subset of extreme CPL-like trajectories are realizable within the underlying two-field dynamics, as visualized in Fig.~\ref{fig:placeholder}. 

Including DESI full-shape information leads to the expected tightening of constraints and a corresponding reduction in the apparent evidence for dynamical dark energy. Within this more constraining likelihood, KMIX remains fully compatible with the data and still allows a mild deviation from $\Lambda$CDM at the $\sim 1.7\sigma$ level. The resulting $(w_0,w_a)$ contours again mirror those from the CPL analysis, but with slightly reduced departure from $(w_0,w_a)=(-1,0)$, reflecting the more restricted space of effective histories that KMIX can generate.

Beyond these cosmological results, a key outcome of this work is methodological. By using conditional normalizing flows to learn the mapping between the high-dimensional KMIX theory space and its effective CPL representation—explicitly accounting for correlations with the background $\Lambda$CDM parameters—we are able to map a realistic, string-motivated model onto the $(w_0,w_a)$ plane and then invert that mapping to obtain direct constraints on the underlying parameters. This strategy enables a direct, like-for-like comparison between physical models and phenomenological fits, while dramatically reducing the computational cost relative to a full MCMC exploration of the KMIX parameter space. Although developed here for a specific axion--dilaton construction, this framework is general and can be applied to a broad class of multi-field dark-energy models or modified-gravity scenarios, provided their expansion histories are sufficiently captured by the CPL parametrization.

Taken together, our results suggest a conservative yet intriguing picture. Current late-time data do not require truly phantom dark energy; instead, they remain fully compatible with a non-phantom theory in which kinetic mixing produces an effective phantom signature when viewed through the CPL lens. Future data---including upcoming DESI
and DESI-II measurements, improved CMB measurements, and next-generation supernova and lensing surveys---will determine whether this ``phantom illusion'' persists. In particular, high-volume surveys such as Spec-S5/MegaMapper will be able to further distinguish these models by constraining the unique signatures from differences in the evolution of perturbations and their imprints on the linear matter power spectrum. Should the preference for evolving dark energy strengthen, models such as KMIX provide a natural and theoretically consistent arena in which to interpret it.

\section*{Acknowledgments}
We thank Stephon Alexander, Heliudson Bernardo, and
Justin Khoury for useful discussions.
We thank Anton Chudaykin 
for sharing with us 
DESI 
MCMC chains
from the full-shape 
likelihood.
This material is based upon work supported by the U.S. Department of Energy, Office of Science, Office of High Energy Physics of U.S. Department of Energy under grant Contract Number  DE-SC0012567. M.W.T.  acknowledges financial support from the Simons Foundation (Grant Number 929255).
JMS acknowledges that support for this work was provided by The Brinson Foundation through a Brinson Prize.

\bibliography{bibo}

\begin{thebibliography}{109}%
\makeatletter
\providecommand \@ifxundefined [1]{%
 \@ifx{#1\undefined}
}%
\providecommand \@ifnum [1]{%
 \ifnum #1\expandafter \@firstoftwo
 \else \expandafter \@secondoftwo
 \fi
}%
\providecommand \@ifx [1]{%
 \ifx #1\expandafter \@firstoftwo
 \else \expandafter \@secondoftwo
 \fi
}%
\providecommand \natexlab [1]{#1}%
\providecommand \enquote  [1]{``#1''}%
\providecommand \bibnamefont  [1]{#1}%
\providecommand \bibfnamefont [1]{#1}%
\providecommand \citenamefont [1]{#1}%
\providecommand \href@noop [0]{\@secondoftwo}%
\providecommand \href [0]{\begingroup \@sanitize@url \@href}%
\providecommand \@href[1]{\@@startlink{#1}\@@href}%
\providecommand \@@href[1]{\endgroup#1\@@endlink}%
\providecommand \@sanitize@url [0]{\catcode `\\12\catcode `\$12\catcode `\&12\catcode `\#12\catcode `\^12\catcode `\_12\catcode `\%12\relax}%
\providecommand \@@startlink[1]{}%
\providecommand \@@endlink[0]{}%
\providecommand \url  [0]{\begingroup\@sanitize@url \@url }%
\providecommand \@url [1]{\endgroup\@href {#1}{\urlprefix }}%
\providecommand \urlprefix  [0]{URL }%
\providecommand \Eprint [0]{\href }%
\providecommand \doibase [0]{http://dx.doi.org/}%
\providecommand \selectlanguage [0]{\@gobble}%
\providecommand \bibinfo  [0]{\@secondoftwo}%
\providecommand \bibfield  [0]{\@secondoftwo}%
\providecommand \translation [1]{[#1]}%
\providecommand \BibitemOpen [0]{}%
\providecommand \bibitemStop [0]{}%
\providecommand \bibitemNoStop [0]{.\EOS\space}%
\providecommand \EOS [0]{\spacefactor3000\relax}%
\providecommand \BibitemShut  [1]{\csname bibitem#1\endcsname}%
\let\auto@bib@innerbib\@empty
\bibitem [{\citenamefont {Adame}\ \emph {et~al.}(2025{\natexlab{a}})\citenamefont {Adame} \emph {et~al.}}]{DESI:2024mwx}%
  \BibitemOpen
  \bibfield  {author} {\bibinfo {author} {\bibfnamefont {A.~G.}\ \bibnamefont {Adame}} \emph {et~al.} (\bibinfo {collaboration} {DESI}),\ }\href {\doibase 10.1088/1475-7516/2025/02/021} {\bibfield  {journal} {\bibinfo  {journal} {JCAP}\ }\textbf {\bibinfo {volume} {02}},\ \bibinfo {pages} {021} (\bibinfo {year} {2025}{\natexlab{a}})},\ \Eprint {http://arxiv.org/abs/2404.03002} {arXiv:2404.03002 [astro-ph.CO]} \BibitemShut {NoStop}%
\bibitem [{\citenamefont {Abdul~Karim}\ \emph {et~al.}(2025)\citenamefont {Abdul~Karim} \emph {et~al.}}]{DESI:2025zgx}%
  \BibitemOpen
  \bibfield  {author} {\bibinfo {author} {\bibfnamefont {M.}~\bibnamefont {Abdul~Karim}} \emph {et~al.} (\bibinfo {collaboration} {DESI}),\ }\href {\doibase 10.1103/tr6y-kpc6} {\bibfield  {journal} {\bibinfo  {journal} {Phys. Rev. D}\ }\textbf {\bibinfo {volume} {112}},\ \bibinfo {pages} {083515} (\bibinfo {year} {2025})},\ \Eprint {http://arxiv.org/abs/2503.14738} {arXiv:2503.14738 [astro-ph.CO]} \BibitemShut {NoStop}%
\bibitem [{\citenamefont {Chevallier}\ and\ \citenamefont {Polarski}(2001)}]{Chevallier:2000qy}%
  \BibitemOpen
  \bibfield  {author} {\bibinfo {author} {\bibfnamefont {M.}~\bibnamefont {Chevallier}}\ and\ \bibinfo {author} {\bibfnamefont {D.}~\bibnamefont {Polarski}},\ }\href {\doibase 10.1142/S0218271801000822} {\bibfield  {journal} {\bibinfo  {journal} {Int. J. Mod. Phys. D}\ }\textbf {\bibinfo {volume} {10}},\ \bibinfo {pages} {213} (\bibinfo {year} {2001})},\ \Eprint {http://arxiv.org/abs/gr-qc/0009008} {arXiv:gr-qc/0009008} \BibitemShut {NoStop}%
\bibitem [{\citenamefont {Linder}(2003)}]{Linder:2002et}%
  \BibitemOpen
  \bibfield  {author} {\bibinfo {author} {\bibfnamefont {E.~V.}\ \bibnamefont {Linder}},\ }\href {\doibase 10.1103/PhysRevLett.90.091301} {\bibfield  {journal} {\bibinfo  {journal} {Phys. Rev. Lett.}\ }\textbf {\bibinfo {volume} {90}},\ \bibinfo {pages} {091301} (\bibinfo {year} {2003})},\ \Eprint {http://arxiv.org/abs/astro-ph/0208512} {arXiv:astro-ph/0208512} \BibitemShut {NoStop}%
\bibitem [{\citenamefont {Lodha}\ \emph {et~al.}(2025{\natexlab{a}})\citenamefont {Lodha} \emph {et~al.}}]{DESI:2025fii}%
  \BibitemOpen
  \bibfield  {author} {\bibinfo {author} {\bibfnamefont {K.}~\bibnamefont {Lodha}} \emph {et~al.} (\bibinfo {collaboration} {DESI}),\ }\href {\doibase 10.1103/w4c6-1r5j} {\bibfield  {journal} {\bibinfo  {journal} {Phys. Rev. D}\ }\textbf {\bibinfo {volume} {112}},\ \bibinfo {pages} {083511} (\bibinfo {year} {2025}{\natexlab{a}})},\ \Eprint {http://arxiv.org/abs/2503.14743} {arXiv:2503.14743 [astro-ph.CO]} \BibitemShut {NoStop}%
\bibitem [{\citenamefont {Chen}\ \emph {et~al.}(2024)\citenamefont {Chen}, \citenamefont {Ivanov}, \citenamefont {Philcox},\ and\ \citenamefont {Wenzl}}]{Chen:2024vuf}%
  \BibitemOpen
  \bibfield  {author} {\bibinfo {author} {\bibfnamefont {S.-F.}\ \bibnamefont {Chen}}, \bibinfo {author} {\bibfnamefont {M.~M.}\ \bibnamefont {Ivanov}}, \bibinfo {author} {\bibfnamefont {O.~H.~E.}\ \bibnamefont {Philcox}}, \ and\ \bibinfo {author} {\bibfnamefont {L.}~\bibnamefont {Wenzl}},\ }\href {\doibase 10.1103/PhysRevLett.133.231001} {\bibfield  {journal} {\bibinfo  {journal} {Phys. Rev. Lett.}\ }\textbf {\bibinfo {volume} {133}},\ \bibinfo {pages} {231001} (\bibinfo {year} {2024})},\ \Eprint {http://arxiv.org/abs/2406.13388} {arXiv:2406.13388 [astro-ph.CO]} \BibitemShut {NoStop}%
\bibitem [{\citenamefont {Lodha}\ \emph {et~al.}(2025{\natexlab{b}})\citenamefont {Lodha} \emph {et~al.}}]{DESI:2024kob}%
  \BibitemOpen
  \bibfield  {author} {\bibinfo {author} {\bibfnamefont {K.}~\bibnamefont {Lodha}} \emph {et~al.} (\bibinfo {collaboration} {DESI}),\ }\href {\doibase 10.1103/PhysRevD.111.023532} {\bibfield  {journal} {\bibinfo  {journal} {Phys. Rev. D}\ }\textbf {\bibinfo {volume} {111}},\ \bibinfo {pages} {023532} (\bibinfo {year} {2025}{\natexlab{b}})},\ \Eprint {http://arxiv.org/abs/2405.13588} {arXiv:2405.13588 [astro-ph.CO]} \BibitemShut {NoStop}%
\bibitem [{\citenamefont {Luongo}\ and\ \citenamefont {Muccino}(2024)}]{Luongo:2024fww}%
  \BibitemOpen
  \bibfield  {author} {\bibinfo {author} {\bibfnamefont {O.}~\bibnamefont {Luongo}}\ and\ \bibinfo {author} {\bibfnamefont {M.}~\bibnamefont {Muccino}},\ }\href {\doibase 10.1051/0004-6361/202450512} {\bibfield  {journal} {\bibinfo  {journal} {Astron. Astrophys.}\ }\textbf {\bibinfo {volume} {690}},\ \bibinfo {pages} {A40} (\bibinfo {year} {2024})},\ \Eprint {http://arxiv.org/abs/2404.07070} {arXiv:2404.07070 [astro-ph.CO]} \BibitemShut {NoStop}%
\bibitem [{\citenamefont {Shajib}\ and\ \citenamefont {Frieman}(2025)}]{Shajib:2025tpd}%
  \BibitemOpen
  \bibfield  {author} {\bibinfo {author} {\bibfnamefont {A.~J.}\ \bibnamefont {Shajib}}\ and\ \bibinfo {author} {\bibfnamefont {J.~A.}\ \bibnamefont {Frieman}},\ }\href {\doibase 10.1103/kjpb-r698} {\bibfield  {journal} {\bibinfo  {journal} {Phys. Rev. D}\ }\textbf {\bibinfo {volume} {112}},\ \bibinfo {pages} {063508} (\bibinfo {year} {2025})},\ \Eprint {http://arxiv.org/abs/2502.06929} {arXiv:2502.06929 [astro-ph.CO]} \BibitemShut {NoStop}%
\bibitem [{\citenamefont {Jiang}\ \emph {et~al.}(2024)\citenamefont {Jiang}, \citenamefont {Pedrotti}, \citenamefont {da~Costa},\ and\ \citenamefont {Vagnozzi}}]{Jiang:2024xnu}%
  \BibitemOpen
  \bibfield  {author} {\bibinfo {author} {\bibfnamefont {J.-Q.}\ \bibnamefont {Jiang}}, \bibinfo {author} {\bibfnamefont {D.}~\bibnamefont {Pedrotti}}, \bibinfo {author} {\bibfnamefont {S.~S.}\ \bibnamefont {da~Costa}}, \ and\ \bibinfo {author} {\bibfnamefont {S.}~\bibnamefont {Vagnozzi}},\ }\href {\doibase 10.1103/PhysRevD.110.123519} {\bibfield  {journal} {\bibinfo  {journal} {Phys. Rev. D}\ }\textbf {\bibinfo {volume} {110}},\ \bibinfo {pages} {123519} (\bibinfo {year} {2024})},\ \Eprint {http://arxiv.org/abs/2408.02365} {arXiv:2408.02365 [astro-ph.CO]} \BibitemShut {NoStop}%
\bibitem [{\citenamefont {Shlivko}\ \emph {et~al.}(2025)\citenamefont {Shlivko}, \citenamefont {Steinhardt},\ and\ \citenamefont {Steinhardt}}]{Shlivko:2025fgv}%
  \BibitemOpen
  \bibfield  {author} {\bibinfo {author} {\bibfnamefont {D.}~\bibnamefont {Shlivko}}, \bibinfo {author} {\bibfnamefont {P.~J.}\ \bibnamefont {Steinhardt}}, \ and\ \bibinfo {author} {\bibfnamefont {C.~L.}\ \bibnamefont {Steinhardt}},\ }\href {\doibase 10.1088/1475-7516/2025/06/054} {\bibfield  {journal} {\bibinfo  {journal} {JCAP}\ }\textbf {\bibinfo {volume} {06}},\ \bibinfo {pages} {054} (\bibinfo {year} {2025})},\ \Eprint {http://arxiv.org/abs/2504.02028} {arXiv:2504.02028 [astro-ph.CO]} \BibitemShut {NoStop}%
\bibitem [{\citenamefont {Efstathiou}(2025)}]{Efstathiou:2025tie}%
  \BibitemOpen
  \bibfield  {author} {\bibinfo {author} {\bibfnamefont {G.}~\bibnamefont {Efstathiou}},\ }\href {\doibase 10.1093/mnras/staf906} {\bibfield  {journal} {\bibinfo  {journal} {Mon. Not. Roy. Astron. Soc.}\ }\textbf {\bibinfo {volume} {540}},\ \bibinfo {pages} {2844} (\bibinfo {year} {2025})},\ \Eprint {http://arxiv.org/abs/2505.02658} {arXiv:2505.02658 [astro-ph.CO]} \BibitemShut {NoStop}%
\bibitem [{\citenamefont {Wang}\ and\ \citenamefont {Freese}(2025)}]{Wang:2025vfb}%
  \BibitemOpen
  \bibfield  {author} {\bibinfo {author} {\bibfnamefont {Y.}~\bibnamefont {Wang}}\ and\ \bibinfo {author} {\bibfnamefont {K.}~\bibnamefont {Freese}},\ }\href@noop {} {\  (\bibinfo {year} {2025})},\ \Eprint {http://arxiv.org/abs/2505.17415} {arXiv:2505.17415 [astro-ph.CO]} \BibitemShut {NoStop}%
\bibitem [{\citenamefont {Cheng}\ \emph {et~al.}(2025)\citenamefont {Cheng}, \citenamefont {Di~Valentino}, \citenamefont {Escamilla}, \citenamefont {Sen},\ and\ \citenamefont {Visinelli}}]{Cheng:2025lod}%
  \BibitemOpen
  \bibfield  {author} {\bibinfo {author} {\bibfnamefont {H.}~\bibnamefont {Cheng}}, \bibinfo {author} {\bibfnamefont {E.}~\bibnamefont {Di~Valentino}}, \bibinfo {author} {\bibfnamefont {L.~A.}\ \bibnamefont {Escamilla}}, \bibinfo {author} {\bibfnamefont {A.~A.}\ \bibnamefont {Sen}}, \ and\ \bibinfo {author} {\bibfnamefont {L.}~\bibnamefont {Visinelli}},\ }\href {\doibase 10.1088/1475-7516/2025/09/031} {\bibfield  {journal} {\bibinfo  {journal} {JCAP}\ }\textbf {\bibinfo {volume} {09}},\ \bibinfo {pages} {031} (\bibinfo {year} {2025})},\ \Eprint {http://arxiv.org/abs/2505.02932} {arXiv:2505.02932 [astro-ph.CO]} \BibitemShut {NoStop}%
\bibitem [{\citenamefont {Chudaykin}\ \emph {et~al.}(2025)\citenamefont {Chudaykin}, \citenamefont {Ivanov},\ and\ \citenamefont {Philcox}}]{Chudaykin:2025aux}%
  \BibitemOpen
  \bibfield  {author} {\bibinfo {author} {\bibfnamefont {A.}~\bibnamefont {Chudaykin}}, \bibinfo {author} {\bibfnamefont {M.~M.}\ \bibnamefont {Ivanov}}, \ and\ \bibinfo {author} {\bibfnamefont {O.~H.~E.}\ \bibnamefont {Philcox}},\ }\href@noop {} {\  (\bibinfo {year} {2025})},\ \Eprint {http://arxiv.org/abs/2507.13433} {arXiv:2507.13433 [astro-ph.CO]} \BibitemShut {NoStop}%
\bibitem [{\citenamefont {de~Souza}\ \emph {et~al.}(2025)\citenamefont {de~Souza}, \citenamefont {Sousa-Neto}, \citenamefont {Gonz{\'a}lez},\ and\ \citenamefont {Alcaniz}}]{deSouza:2025vdv}%
  \BibitemOpen
  \bibfield  {author} {\bibinfo {author} {\bibfnamefont {R.}~\bibnamefont {de~Souza}}, \bibinfo {author} {\bibfnamefont {A.}~\bibnamefont {Sousa-Neto}}, \bibinfo {author} {\bibfnamefont {J.~E.}\ \bibnamefont {Gonz{\'a}lez}}, \ and\ \bibinfo {author} {\bibfnamefont {J.}~\bibnamefont {Alcaniz}},\ }\href@noop {} {\  (\bibinfo {year} {2025})},\ \Eprint {http://arxiv.org/abs/2511.13666} {arXiv:2511.13666 [astro-ph.CO]} \BibitemShut {NoStop}%
\bibitem [{\citenamefont {Toomey}\ \emph {et~al.}(2025{\natexlab{a}})\citenamefont {Toomey}, \citenamefont {Montefalcone}, \citenamefont {McDonough},\ and\ \citenamefont {Freese}}]{Toomey:2025xyo}%
  \BibitemOpen
  \bibfield  {author} {\bibinfo {author} {\bibfnamefont {M.~W.}\ \bibnamefont {Toomey}}, \bibinfo {author} {\bibfnamefont {G.}~\bibnamefont {Montefalcone}}, \bibinfo {author} {\bibfnamefont {E.}~\bibnamefont {McDonough}}, \ and\ \bibinfo {author} {\bibfnamefont {K.}~\bibnamefont {Freese}},\ }\href@noop {} {\  (\bibinfo {year} {2025}{\natexlab{a}})},\ \Eprint {http://arxiv.org/abs/2509.13318} {arXiv:2509.13318 [astro-ph.CO]} \BibitemShut {NoStop}%
\bibitem [{\citenamefont {Ong}\ \emph {et~al.}(2025)\citenamefont {Ong}, \citenamefont {Yallup},\ and\ \citenamefont {Handley}}]{Ong:2025utx}%
  \BibitemOpen
  \bibfield  {author} {\bibinfo {author} {\bibfnamefont {D.~D.~Y.}\ \bibnamefont {Ong}}, \bibinfo {author} {\bibfnamefont {D.}~\bibnamefont {Yallup}}, \ and\ \bibinfo {author} {\bibfnamefont {W.}~\bibnamefont {Handley}},\ }\href@noop {} {\  (\bibinfo {year} {2025})},\ \Eprint {http://arxiv.org/abs/2511.10631} {arXiv:2511.10631 [astro-ph.CO]} \BibitemShut {NoStop}%
\bibitem [{\citenamefont {Caldwell}(2002)}]{Caldwell:1999ew}%
  \BibitemOpen
  \bibfield  {author} {\bibinfo {author} {\bibfnamefont {R.~R.}\ \bibnamefont {Caldwell}},\ }\href {\doibase 10.1016/S0370-2693(02)02589-3} {\bibfield  {journal} {\bibinfo  {journal} {Phys. Lett. B}\ }\textbf {\bibinfo {volume} {545}},\ \bibinfo {pages} {23} (\bibinfo {year} {2002})},\ \Eprint {http://arxiv.org/abs/astro-ph/9908168} {arXiv:astro-ph/9908168} \BibitemShut {NoStop}%
\bibitem [{\citenamefont {Hu}(2005)}]{Hu:2004kh}%
  \BibitemOpen
  \bibfield  {author} {\bibinfo {author} {\bibfnamefont {W.}~\bibnamefont {Hu}},\ }\href {\doibase 10.1103/PhysRevD.71.047301} {\bibfield  {journal} {\bibinfo  {journal} {Phys. Rev. D}\ }\textbf {\bibinfo {volume} {71}},\ \bibinfo {pages} {047301} (\bibinfo {year} {2005})},\ \Eprint {http://arxiv.org/abs/astro-ph/0410680} {arXiv:astro-ph/0410680} \BibitemShut {NoStop}%
\bibitem [{\citenamefont {Caldwell}\ and\ \citenamefont {Linder}(2025)}]{Caldwell:2025inn}%
  \BibitemOpen
  \bibfield  {author} {\bibinfo {author} {\bibfnamefont {R.~R.}\ \bibnamefont {Caldwell}}\ and\ \bibinfo {author} {\bibfnamefont {E.~V.}\ \bibnamefont {Linder}},\ }\href@noop {} {\  (\bibinfo {year} {2025})},\ \Eprint {http://arxiv.org/abs/2511.07526} {arXiv:2511.07526 [astro-ph.CO]} \BibitemShut {NoStop}%
\bibitem [{\citenamefont {Das}\ \emph {et~al.}(2006)\citenamefont {Das}, \citenamefont {Corasaniti},\ and\ \citenamefont {Khoury}}]{Das:2005yj}%
  \BibitemOpen
  \bibfield  {author} {\bibinfo {author} {\bibfnamefont {S.}~\bibnamefont {Das}}, \bibinfo {author} {\bibfnamefont {P.~S.}\ \bibnamefont {Corasaniti}}, \ and\ \bibinfo {author} {\bibfnamefont {J.}~\bibnamefont {Khoury}},\ }\href {\doibase 10.1103/PhysRevD.73.083509} {\bibfield  {journal} {\bibinfo  {journal} {Phys. Rev. D}\ }\textbf {\bibinfo {volume} {73}},\ \bibinfo {pages} {083509} (\bibinfo {year} {2006})},\ \Eprint {http://arxiv.org/abs/astro-ph/0510628} {arXiv:astro-ph/0510628} \BibitemShut {NoStop}%
\bibitem [{\citenamefont {Carroll}\ \emph {et~al.}(2005)\citenamefont {Carroll}, \citenamefont {De~Felice},\ and\ \citenamefont {Trodden}}]{Carroll:2004hc}%
  \BibitemOpen
  \bibfield  {author} {\bibinfo {author} {\bibfnamefont {S.~M.}\ \bibnamefont {Carroll}}, \bibinfo {author} {\bibfnamefont {A.}~\bibnamefont {De~Felice}}, \ and\ \bibinfo {author} {\bibfnamefont {M.}~\bibnamefont {Trodden}},\ }\href {\doibase 10.1103/PhysRevD.71.023525} {\bibfield  {journal} {\bibinfo  {journal} {Phys. Rev. D}\ }\textbf {\bibinfo {volume} {71}},\ \bibinfo {pages} {023525} (\bibinfo {year} {2005})},\ \Eprint {http://arxiv.org/abs/astro-ph/0408081} {arXiv:astro-ph/0408081} \BibitemShut {NoStop}%
\bibitem [{\citenamefont {Khoury}\ \emph {et~al.}(2025)\citenamefont {Khoury}, \citenamefont {Lin},\ and\ \citenamefont {Trodden}}]{Khoury:2025txd}%
  \BibitemOpen
  \bibfield  {author} {\bibinfo {author} {\bibfnamefont {J.}~\bibnamefont {Khoury}}, \bibinfo {author} {\bibfnamefont {M.-X.}\ \bibnamefont {Lin}}, \ and\ \bibinfo {author} {\bibfnamefont {M.}~\bibnamefont {Trodden}},\ }\href {\doibase 10.1103/w4qb-plk8} {\bibfield  {journal} {\bibinfo  {journal} {Phys. Rev. Lett.}\ }\textbf {\bibinfo {volume} {135}},\ \bibinfo {pages} {181001} (\bibinfo {year} {2025})},\ \Eprint {http://arxiv.org/abs/2503.16415} {arXiv:2503.16415 [astro-ph.CO]} \BibitemShut {NoStop}%
\bibitem [{\citenamefont {Huey}\ and\ \citenamefont {Wandelt}(2006)}]{Huey:2004qv}%
  \BibitemOpen
  \bibfield  {author} {\bibinfo {author} {\bibfnamefont {G.}~\bibnamefont {Huey}}\ and\ \bibinfo {author} {\bibfnamefont {B.~D.}\ \bibnamefont {Wandelt}},\ }\href {\doibase 10.1103/PhysRevD.74.023519} {\bibfield  {journal} {\bibinfo  {journal} {Phys. Rev. D}\ }\textbf {\bibinfo {volume} {74}},\ \bibinfo {pages} {023519} (\bibinfo {year} {2006})},\ \Eprint {http://arxiv.org/abs/astro-ph/0407196} {arXiv:astro-ph/0407196} \BibitemShut {NoStop}%
\bibitem [{\citenamefont {Brax}\ \emph {et~al.}(2024)\citenamefont {Brax}, \citenamefont {Burgess},\ and\ \citenamefont {Quevedo}}]{Brax:2023qyp}%
  \BibitemOpen
  \bibfield  {author} {\bibinfo {author} {\bibfnamefont {P.}~\bibnamefont {Brax}}, \bibinfo {author} {\bibfnamefont {C.~P.}\ \bibnamefont {Burgess}}, \ and\ \bibinfo {author} {\bibfnamefont {F.}~\bibnamefont {Quevedo}},\ }\href {\doibase 10.1088/1475-7516/2024/03/015} {\bibfield  {journal} {\bibinfo  {journal} {JCAP}\ }\textbf {\bibinfo {volume} {03}},\ \bibinfo {pages} {015} (\bibinfo {year} {2024})},\ \Eprint {http://arxiv.org/abs/2310.02092} {arXiv:2310.02092 [hep-th]} \BibitemShut {NoStop}%
\bibitem [{\citenamefont {Smith}\ \emph {et~al.}(2025)\citenamefont {Smith}, \citenamefont {Mylova}, \citenamefont {Brax}, \citenamefont {van~de Bruck}, \citenamefont {Burgess},\ and\ \citenamefont {Davis}}]{Smith:2024ibv}%
  \BibitemOpen
  \bibfield  {author} {\bibinfo {author} {\bibfnamefont {A.}~\bibnamefont {Smith}}, \bibinfo {author} {\bibfnamefont {M.}~\bibnamefont {Mylova}}, \bibinfo {author} {\bibfnamefont {P.}~\bibnamefont {Brax}}, \bibinfo {author} {\bibfnamefont {C.}~\bibnamefont {van~de Bruck}}, \bibinfo {author} {\bibfnamefont {C.~P.}\ \bibnamefont {Burgess}}, \ and\ \bibinfo {author} {\bibfnamefont {A.-C.}\ \bibnamefont {Davis}},\ }\href {\doibase 10.1088/1475-7516/2025/07/023} {\bibfield  {journal} {\bibinfo  {journal} {JCAP}\ }\textbf {\bibinfo {volume} {07}},\ \bibinfo {pages} {023} (\bibinfo {year} {2025})},\ \Eprint {http://arxiv.org/abs/2410.11099} {arXiv:2410.11099 [hep-th]} \BibitemShut {NoStop}%
\bibitem [{\citenamefont {Bedroya}\ \emph {et~al.}(2025)\citenamefont {Bedroya}, \citenamefont {Obied}, \citenamefont {Vafa},\ and\ \citenamefont {Wu}}]{Bedroya:2025fwh}%
  \BibitemOpen
  \bibfield  {author} {\bibinfo {author} {\bibfnamefont {A.}~\bibnamefont {Bedroya}}, \bibinfo {author} {\bibfnamefont {G.}~\bibnamefont {Obied}}, \bibinfo {author} {\bibfnamefont {C.}~\bibnamefont {Vafa}}, \ and\ \bibinfo {author} {\bibfnamefont {D.~H.}\ \bibnamefont {Wu}},\ }\href@noop {} {\  (\bibinfo {year} {2025})},\ \Eprint {http://arxiv.org/abs/2507.03090} {arXiv:2507.03090 [astro-ph.CO]} \BibitemShut {NoStop}%
\bibitem [{\citenamefont {Silva}\ \emph {et~al.}(2025)\citenamefont {Silva}, \citenamefont {Sabogal}, \citenamefont {Scherer}, \citenamefont {Nunes}, \citenamefont {Di~Valentino},\ and\ \citenamefont {Kumar}}]{Silva:2025hxw}%
  \BibitemOpen
  \bibfield  {author} {\bibinfo {author} {\bibfnamefont {E.}~\bibnamefont {Silva}}, \bibinfo {author} {\bibfnamefont {M.~A.}\ \bibnamefont {Sabogal}}, \bibinfo {author} {\bibfnamefont {M.}~\bibnamefont {Scherer}}, \bibinfo {author} {\bibfnamefont {R.~C.}\ \bibnamefont {Nunes}}, \bibinfo {author} {\bibfnamefont {E.}~\bibnamefont {Di~Valentino}}, \ and\ \bibinfo {author} {\bibfnamefont {S.}~\bibnamefont {Kumar}},\ }\href {\doibase 10.1103/qqc6-76z4} {\bibfield  {journal} {\bibinfo  {journal} {Phys. Rev. D}\ }\textbf {\bibinfo {volume} {111}},\ \bibinfo {pages} {123511} (\bibinfo {year} {2025})},\ \Eprint {http://arxiv.org/abs/2503.23225} {arXiv:2503.23225 [astro-ph.CO]} \BibitemShut {NoStop}%
\bibitem [{\citenamefont {Wolf}\ \emph {et~al.}(2024)\citenamefont {Wolf}, \citenamefont {Garc{\'\i}a-Garc{\'\i}a}, \citenamefont {Bartlett},\ and\ \citenamefont {Ferreira}}]{Wolf:2024eph}%
  \BibitemOpen
  \bibfield  {author} {\bibinfo {author} {\bibfnamefont {W.~J.}\ \bibnamefont {Wolf}}, \bibinfo {author} {\bibfnamefont {C.}~\bibnamefont {Garc{\'\i}a-Garc{\'\i}a}}, \bibinfo {author} {\bibfnamefont {D.~J.}\ \bibnamefont {Bartlett}}, \ and\ \bibinfo {author} {\bibfnamefont {P.~G.}\ \bibnamefont {Ferreira}},\ }\href {\doibase 10.1103/PhysRevD.110.083528} {\bibfield  {journal} {\bibinfo  {journal} {Phys. Rev. D}\ }\textbf {\bibinfo {volume} {110}},\ \bibinfo {pages} {083528} (\bibinfo {year} {2024})},\ \Eprint {http://arxiv.org/abs/2408.17318} {arXiv:2408.17318 [astro-ph.CO]} \BibitemShut {NoStop}%
\bibitem [{\citenamefont {Wolf}\ \emph {et~al.}(2025{\natexlab{a}})\citenamefont {Wolf}, \citenamefont {Ferreira},\ and\ \citenamefont {Garc{\'\i}a-Garc{\'\i}a}}]{Wolf:2024stt}%
  \BibitemOpen
  \bibfield  {author} {\bibinfo {author} {\bibfnamefont {W.~J.}\ \bibnamefont {Wolf}}, \bibinfo {author} {\bibfnamefont {P.~G.}\ \bibnamefont {Ferreira}}, \ and\ \bibinfo {author} {\bibfnamefont {C.}~\bibnamefont {Garc{\'\i}a-Garc{\'\i}a}},\ }\href {\doibase 10.1103/PhysRevD.111.L041303} {\bibfield  {journal} {\bibinfo  {journal} {Phys. Rev. D}\ }\textbf {\bibinfo {volume} {111}},\ \bibinfo {pages} {L041303} (\bibinfo {year} {2025}{\natexlab{a}})},\ \Eprint {http://arxiv.org/abs/2409.17019} {arXiv:2409.17019 [astro-ph.CO]} \BibitemShut {NoStop}%
\bibitem [{\citenamefont {Wolf}\ \emph {et~al.}(2025{\natexlab{b}})\citenamefont {Wolf}, \citenamefont {Garc{\'\i}a-Garc{\'\i}a}, \citenamefont {Anton},\ and\ \citenamefont {Ferreira}}]{Wolf:2025jed}%
  \BibitemOpen
  \bibfield  {author} {\bibinfo {author} {\bibfnamefont {W.~J.}\ \bibnamefont {Wolf}}, \bibinfo {author} {\bibfnamefont {C.}~\bibnamefont {Garc{\'\i}a-Garc{\'\i}a}}, \bibinfo {author} {\bibfnamefont {T.}~\bibnamefont {Anton}}, \ and\ \bibinfo {author} {\bibfnamefont {P.~G.}\ \bibnamefont {Ferreira}},\ }\href {\doibase 10.1103/jysf-k72m} {\bibfield  {journal} {\bibinfo  {journal} {Phys. Rev. Lett.}\ }\textbf {\bibinfo {volume} {135}},\ \bibinfo {pages} {081001} (\bibinfo {year} {2025}{\natexlab{b}})},\ \Eprint {http://arxiv.org/abs/2504.07679} {arXiv:2504.07679 [astro-ph.CO]} \BibitemShut {NoStop}%
\bibitem [{\citenamefont {Goldstein}\ \emph {et~al.}(2025)\citenamefont {Goldstein}, \citenamefont {Celoria},\ and\ \citenamefont {Schmidt}}]{Goldstein:2025epp}%
  \BibitemOpen
  \bibfield  {author} {\bibinfo {author} {\bibfnamefont {S.}~\bibnamefont {Goldstein}}, \bibinfo {author} {\bibfnamefont {M.}~\bibnamefont {Celoria}}, \ and\ \bibinfo {author} {\bibfnamefont {F.}~\bibnamefont {Schmidt}},\ }\href@noop {} {\  (\bibinfo {year} {2025})},\ \Eprint {http://arxiv.org/abs/2507.16970} {arXiv:2507.16970 [astro-ph.CO]} \BibitemShut {NoStop}%
\bibitem [{\citenamefont {Bottaro}\ \emph {et~al.}(2025)\citenamefont {Bottaro}, \citenamefont {Castorina}, \citenamefont {Costa}, \citenamefont {Redigolo},\ and\ \citenamefont {Salvioni}}]{Bottaro:2024pcb}%
  \BibitemOpen
  \bibfield  {author} {\bibinfo {author} {\bibfnamefont {S.}~\bibnamefont {Bottaro}}, \bibinfo {author} {\bibfnamefont {E.}~\bibnamefont {Castorina}}, \bibinfo {author} {\bibfnamefont {M.}~\bibnamefont {Costa}}, \bibinfo {author} {\bibfnamefont {D.}~\bibnamefont {Redigolo}}, \ and\ \bibinfo {author} {\bibfnamefont {E.}~\bibnamefont {Salvioni}},\ }\href {\doibase 10.1103/gc78-96l5} {\bibfield  {journal} {\bibinfo  {journal} {Phys. Rev. D}\ }\textbf {\bibinfo {volume} {112}},\ \bibinfo {pages} {023525} (\bibinfo {year} {2025})},\ \Eprint {http://arxiv.org/abs/2407.18252} {arXiv:2407.18252 [astro-ph.CO]} \BibitemShut {NoStop}%
\bibitem [{\citenamefont {Burgess}(2025)}]{Burgess:2025vxs}%
  \BibitemOpen
  \bibfield  {author} {\bibinfo {author} {\bibfnamefont {C.~P.}\ \bibnamefont {Burgess}}\ }(\bibinfo {year} {2025})\ \Eprint {http://arxiv.org/abs/2509.00688} {arXiv:2509.00688 [hep-th]} \BibitemShut {NoStop}%
\bibitem [{\citenamefont {Peebles}\ and\ \citenamefont {Ratra}(1988)}]{Peebles:1987ek}%
  \BibitemOpen
  \bibfield  {author} {\bibinfo {author} {\bibfnamefont {P.~J.~E.}\ \bibnamefont {Peebles}}\ and\ \bibinfo {author} {\bibfnamefont {B.}~\bibnamefont {Ratra}},\ }\href {\doibase 10.1086/185100} {\bibfield  {journal} {\bibinfo  {journal} {Astrophys. J. Lett.}\ }\textbf {\bibinfo {volume} {325}},\ \bibinfo {pages} {L17} (\bibinfo {year} {1988})}\BibitemShut {NoStop}%
\bibitem [{\citenamefont {Ratra}\ and\ \citenamefont {Peebles}(1988)}]{Ratra:1987rm}%
  \BibitemOpen
  \bibfield  {author} {\bibinfo {author} {\bibfnamefont {B.}~\bibnamefont {Ratra}}\ and\ \bibinfo {author} {\bibfnamefont {P.~J.~E.}\ \bibnamefont {Peebles}},\ }\href {\doibase 10.1103/PhysRevD.37.3406} {\bibfield  {journal} {\bibinfo  {journal} {Phys. Rev.}\ }\textbf {\bibinfo {volume} {D37}},\ \bibinfo {pages} {3406} (\bibinfo {year} {1988})}\BibitemShut {NoStop}%
\bibitem [{\citenamefont {Bernardo}\ \emph {et~al.}(2022)\citenamefont {Bernardo}, \citenamefont {Brandenberger},\ and\ \citenamefont {Fr\"ohlich}}]{Bernardo:2022ztc}%
  \BibitemOpen
  \bibfield  {author} {\bibinfo {author} {\bibfnamefont {H.}~\bibnamefont {Bernardo}}, \bibinfo {author} {\bibfnamefont {R.}~\bibnamefont {Brandenberger}}, \ and\ \bibinfo {author} {\bibfnamefont {J.}~\bibnamefont {Fr\"ohlich}},\ }\href {\doibase 10.1088/1475-7516/2022/09/040} {\bibfield  {journal} {\bibinfo  {journal} {JCAP}\ }\textbf {\bibinfo {volume} {09}},\ \bibinfo {pages} {040} (\bibinfo {year} {2022})},\ \Eprint {http://arxiv.org/abs/2201.04668} {arXiv:2201.04668 [hep-th]} \BibitemShut {NoStop}%
\bibitem [{\citenamefont {Alexander}\ \emph {et~al.}(2023)\citenamefont {Alexander}, \citenamefont {Bernardo},\ and\ \citenamefont {Toomey}}]{Alexander:2022own}%
  \BibitemOpen
  \bibfield  {author} {\bibinfo {author} {\bibfnamefont {S.}~\bibnamefont {Alexander}}, \bibinfo {author} {\bibfnamefont {H.}~\bibnamefont {Bernardo}}, \ and\ \bibinfo {author} {\bibfnamefont {M.~W.}\ \bibnamefont {Toomey}},\ }\href {\doibase 10.1088/1475-7516/2023/03/037} {\bibfield  {journal} {\bibinfo  {journal} {JCAP}\ }\textbf {\bibinfo {volume} {03}},\ \bibinfo {pages} {037} (\bibinfo {year} {2023})},\ \Eprint {http://arxiv.org/abs/2207.13086} {arXiv:2207.13086 [astro-ph.CO]} \BibitemShut {NoStop}%
\bibitem [{\citenamefont {Toomey}\ \emph {et~al.}(2025{\natexlab{b}})\citenamefont {Toomey}, \citenamefont {Koushiappas},\ and\ \citenamefont {Alexander}}]{Toomey:2025mvx}%
  \BibitemOpen
  \bibfield  {author} {\bibinfo {author} {\bibfnamefont {M.~W.}\ \bibnamefont {Toomey}}, \bibinfo {author} {\bibfnamefont {S.~M.}\ \bibnamefont {Koushiappas}}, \ and\ \bibinfo {author} {\bibfnamefont {S.}~\bibnamefont {Alexander}},\ }\href@noop {} {\  (\bibinfo {year} {2025}{\natexlab{b}})},\ \Eprint {http://arxiv.org/abs/2506.08076} {arXiv:2506.08076 [astro-ph.CO]} \BibitemShut {NoStop}%
\bibitem [{\citenamefont {Toomey}\ \emph {et~al.}(2025{\natexlab{c}})\citenamefont {Toomey}, \citenamefont {Ivanov},\ and\ \citenamefont {McDonough}}]{Toomey:2024ita}%
  \BibitemOpen
  \bibfield  {author} {\bibinfo {author} {\bibfnamefont {M.~W.}\ \bibnamefont {Toomey}}, \bibinfo {author} {\bibfnamefont {M.~M.}\ \bibnamefont {Ivanov}}, \ and\ \bibinfo {author} {\bibfnamefont {E.}~\bibnamefont {McDonough}},\ }\href {\doibase 10.1103/1wr4-x5dy} {\bibfield  {journal} {\bibinfo  {journal} {Phys. Rev. D}\ }\textbf {\bibinfo {volume} {112}},\ \bibinfo {pages} {103539} (\bibinfo {year} {2025}{\natexlab{c}})},\ \Eprint {http://arxiv.org/abs/2409.09029} {arXiv:2409.09029 [astro-ph.CO]} \BibitemShut {NoStop}%
\bibitem [{\citenamefont {Ivanov}\ \emph {et~al.}(2024{\natexlab{a}})\citenamefont {Ivanov}, \citenamefont {Cuesta-Lazaro}, \citenamefont {Mishra-Sharma}, \citenamefont {Obuljen},\ and\ \citenamefont {Toomey}}]{Ivanov:2024hgq}%
  \BibitemOpen
  \bibfield  {author} {\bibinfo {author} {\bibfnamefont {M.~M.}\ \bibnamefont {Ivanov}}, \bibinfo {author} {\bibfnamefont {C.}~\bibnamefont {Cuesta-Lazaro}}, \bibinfo {author} {\bibfnamefont {S.}~\bibnamefont {Mishra-Sharma}}, \bibinfo {author} {\bibfnamefont {A.}~\bibnamefont {Obuljen}}, \ and\ \bibinfo {author} {\bibfnamefont {M.~W.}\ \bibnamefont {Toomey}},\ }\href {\doibase 10.1103/PhysRevD.110.063538} {\bibfield  {journal} {\bibinfo  {journal} {Phys. Rev. D}\ }\textbf {\bibinfo {volume} {110}},\ \bibinfo {pages} {063538} (\bibinfo {year} {2024}{\natexlab{a}})},\ \Eprint {http://arxiv.org/abs/2402.13310} {arXiv:2402.13310 [astro-ph.CO]} \BibitemShut {NoStop}%
\bibitem [{\citenamefont {Ivanov}\ \emph {et~al.}(2025)\citenamefont {Ivanov}, \citenamefont {Obuljen}, \citenamefont {Cuesta-Lazaro},\ and\ \citenamefont {Toomey}}]{Ivanov:2024xgb}%
  \BibitemOpen
  \bibfield  {author} {\bibinfo {author} {\bibfnamefont {M.~M.}\ \bibnamefont {Ivanov}}, \bibinfo {author} {\bibfnamefont {A.}~\bibnamefont {Obuljen}}, \bibinfo {author} {\bibfnamefont {C.}~\bibnamefont {Cuesta-Lazaro}}, \ and\ \bibinfo {author} {\bibfnamefont {M.~W.}\ \bibnamefont {Toomey}},\ }\href {\doibase 10.1103/PhysRevD.111.063548} {\bibfield  {journal} {\bibinfo  {journal} {Phys. Rev. D}\ }\textbf {\bibinfo {volume} {111}},\ \bibinfo {pages} {063548} (\bibinfo {year} {2025})},\ \Eprint {http://arxiv.org/abs/2409.10609} {arXiv:2409.10609 [astro-ph.CO]} \BibitemShut {NoStop}%
\bibitem [{\citenamefont {Ivanov}\ \emph {et~al.}(2024{\natexlab{b}})\citenamefont {Ivanov} \emph {et~al.}}]{Ivanov:2024dgv}%
  \BibitemOpen
  \bibfield  {author} {\bibinfo {author} {\bibfnamefont {M.~M.}\ \bibnamefont {Ivanov}} \emph {et~al.},\ }\href@noop {} {\  (\bibinfo {year} {2024}{\natexlab{b}})},\ \Eprint {http://arxiv.org/abs/2412.01888} {arXiv:2412.01888 [astro-ph.CO]} \BibitemShut {NoStop}%
\bibitem [{\citenamefont {Chen}\ and\ \citenamefont {Ivanov}(2025)}]{Chen:2025jnr}%
  \BibitemOpen
  \bibfield  {author} {\bibinfo {author} {\bibfnamefont {S.-F.}\ \bibnamefont {Chen}}\ and\ \bibinfo {author} {\bibfnamefont {M.~M.}\ \bibnamefont {Ivanov}},\ }\href {\doibase 10.1103/3nbk-l5v1} {\bibfield  {journal} {\bibinfo  {journal} {Phys. Rev. D}\ }\textbf {\bibinfo {volume} {112}},\ \bibinfo {pages} {083562} (\bibinfo {year} {2025})},\ \Eprint {http://arxiv.org/abs/2507.00118} {arXiv:2507.00118 [astro-ph.CO]} \BibitemShut {NoStop}%
\bibitem [{\citenamefont {Cicoli}\ \emph {et~al.}(2012)\citenamefont {Cicoli}, \citenamefont {Goodsell},\ and\ \citenamefont {Ringwald}}]{Cicoli:2012sz}%
  \BibitemOpen
  \bibfield  {author} {\bibinfo {author} {\bibfnamefont {M.}~\bibnamefont {Cicoli}}, \bibinfo {author} {\bibfnamefont {M.}~\bibnamefont {Goodsell}}, \ and\ \bibinfo {author} {\bibfnamefont {A.}~\bibnamefont {Ringwald}},\ }\href {\doibase 10.1007/JHEP10(2012)146} {\bibfield  {journal} {\bibinfo  {journal} {JHEP}\ }\textbf {\bibinfo {volume} {10}},\ \bibinfo {pages} {146} (\bibinfo {year} {2012})},\ \Eprint {http://arxiv.org/abs/1206.0819} {arXiv:1206.0819 [hep-th]} \BibitemShut {NoStop}%
\bibitem [{\citenamefont {Arvanitaki}\ \emph {et~al.}(2010)\citenamefont {Arvanitaki}, \citenamefont {Dimopoulos}, \citenamefont {Dubovsky}, \citenamefont {Kaloper},\ and\ \citenamefont {March-Russell}}]{Arvanitaki:2009fg}%
  \BibitemOpen
  \bibfield  {author} {\bibinfo {author} {\bibfnamefont {A.}~\bibnamefont {Arvanitaki}}, \bibinfo {author} {\bibfnamefont {S.}~\bibnamefont {Dimopoulos}}, \bibinfo {author} {\bibfnamefont {S.}~\bibnamefont {Dubovsky}}, \bibinfo {author} {\bibfnamefont {N.}~\bibnamefont {Kaloper}}, \ and\ \bibinfo {author} {\bibfnamefont {J.}~\bibnamefont {March-Russell}},\ }\href {\doibase 10.1103/PhysRevD.81.123530} {\bibfield  {journal} {\bibinfo  {journal} {Phys. Rev. D}\ }\textbf {\bibinfo {volume} {81}},\ \bibinfo {pages} {123530} (\bibinfo {year} {2010})},\ \Eprint {http://arxiv.org/abs/0905.4720} {arXiv:0905.4720 [hep-th]} \BibitemShut {NoStop}%
\bibitem [{\citenamefont {Burgess}\ \emph {et~al.}(2022)\citenamefont {Burgess}, \citenamefont {Dineen},\ and\ \citenamefont {Quevedo}}]{Burgess:2021obw}%
  \BibitemOpen
  \bibfield  {author} {\bibinfo {author} {\bibfnamefont {C.~P.}\ \bibnamefont {Burgess}}, \bibinfo {author} {\bibfnamefont {D.}~\bibnamefont {Dineen}}, \ and\ \bibinfo {author} {\bibfnamefont {F.}~\bibnamefont {Quevedo}},\ }\href {\doibase 10.1088/1475-7516/2022/03/064} {\bibfield  {journal} {\bibinfo  {journal} {JCAP}\ }\textbf {\bibinfo {volume} {03}},\ \bibinfo {pages} {064} (\bibinfo {year} {2022})},\ \Eprint {http://arxiv.org/abs/2111.07286} {arXiv:2111.07286 [hep-th]} \BibitemShut {NoStop}%
\bibitem [{\citenamefont {Berera}\ \emph {et~al.}(2025)\citenamefont {Berera}, \citenamefont {Bernardo}, \citenamefont {Brahma}, \citenamefont {Calder{\'o}n-Figueroa}, \citenamefont {O.~Ramos},\ and\ \citenamefont {Toomey}}]{Berera:2025jbj}%
  \BibitemOpen
  \bibfield  {author} {\bibinfo {author} {\bibfnamefont {A.}~\bibnamefont {Berera}}, \bibinfo {author} {\bibfnamefont {H.}~\bibnamefont {Bernardo}}, \bibinfo {author} {\bibfnamefont {S.}~\bibnamefont {Brahma}}, \bibinfo {author} {\bibfnamefont {J.}~\bibnamefont {Calder{\'o}n-Figueroa}}, \bibinfo {author} {\bibfnamefont {R.}~\bibnamefont {O.~Ramos}}, \ and\ \bibinfo {author} {\bibfnamefont {M.~W.}\ \bibnamefont {Toomey}},\ }\href@noop {} {\  (\bibinfo {year} {2025})},\ \Eprint {http://arxiv.org/abs/2509.15125} {arXiv:2509.15125 [hep-th]} \BibitemShut {NoStop}%
\bibitem [{\citenamefont {Alexander}\ and\ \citenamefont {McDonough}(2019)}]{Alexander:2019rsc}%
  \BibitemOpen
  \bibfield  {author} {\bibinfo {author} {\bibfnamefont {S.}~\bibnamefont {Alexander}}\ and\ \bibinfo {author} {\bibfnamefont {E.}~\bibnamefont {McDonough}},\ }\href {\doibase 10.1016/j.physletb.2019.134830} {\bibfield  {journal} {\bibinfo  {journal} {Phys. Lett. B}\ }\textbf {\bibinfo {volume} {797}},\ \bibinfo {pages} {134830} (\bibinfo {year} {2019})},\ \Eprint {http://arxiv.org/abs/1904.08912} {arXiv:1904.08912 [astro-ph.CO]} \BibitemShut {NoStop}%
\bibitem [{\citenamefont {{Papamakarios}}\ \emph {et~al.}(2021)\citenamefont {{Papamakarios}}, \citenamefont {{Nalisnick}}, \citenamefont {{Jimenez Rezende}}, \citenamefont {{Mohamed}},\ and\ \citenamefont {{Lakshminarayanan}}}]{2019arXiv191202762P}%
  \BibitemOpen
  \bibfield  {author} {\bibinfo {author} {\bibfnamefont {G.}~\bibnamefont {{Papamakarios}}}, \bibinfo {author} {\bibfnamefont {E.}~\bibnamefont {{Nalisnick}}}, \bibinfo {author} {\bibfnamefont {D.}~\bibnamefont {{Jimenez Rezende}}}, \bibinfo {author} {\bibfnamefont {S.}~\bibnamefont {{Mohamed}}}, \ and\ \bibinfo {author} {\bibfnamefont {B.}~\bibnamefont {{Lakshminarayanan}}},\ }\href {http://jmlr.org/papers/v22/19-1028.html} {\bibfield  {journal} {\bibinfo  {journal} {Journal of Machine Learning Research}\ }\textbf {\bibinfo {volume} {22}},\ \bibinfo {pages} {1} (\bibinfo {year} {2021})},\ \Eprint {http://arxiv.org/abs/1912.02762} {arXiv:1912.02762 [stat.ML]} \BibitemShut {NoStop}%
\bibitem [{\citenamefont {Kallosh}\ \emph {et~al.}(1995)\citenamefont {Kallosh}, \citenamefont {Linde}, \citenamefont {Linde},\ and\ \citenamefont {Susskind}}]{Kallosh:1995hi}%
  \BibitemOpen
  \bibfield  {author} {\bibinfo {author} {\bibfnamefont {R.}~\bibnamefont {Kallosh}}, \bibinfo {author} {\bibfnamefont {A.~D.}\ \bibnamefont {Linde}}, \bibinfo {author} {\bibfnamefont {D.~A.}\ \bibnamefont {Linde}}, \ and\ \bibinfo {author} {\bibfnamefont {L.}~\bibnamefont {Susskind}},\ }\href {\doibase 10.1103/PhysRevD.52.912} {\bibfield  {journal} {\bibinfo  {journal} {Phys. Rev. D}\ }\textbf {\bibinfo {volume} {52}},\ \bibinfo {pages} {912} (\bibinfo {year} {1995})},\ \Eprint {http://arxiv.org/abs/hep-th/9502069} {arXiv:hep-th/9502069} \BibitemShut {NoStop}%
\bibitem [{\citenamefont {Banks}\ \emph {et~al.}(2003)\citenamefont {Banks}, \citenamefont {Dine}, \citenamefont {Fox},\ and\ \citenamefont {Gorbatov}}]{Banks:2003sx}%
  \BibitemOpen
  \bibfield  {author} {\bibinfo {author} {\bibfnamefont {T.}~\bibnamefont {Banks}}, \bibinfo {author} {\bibfnamefont {M.}~\bibnamefont {Dine}}, \bibinfo {author} {\bibfnamefont {P.~J.}\ \bibnamefont {Fox}}, \ and\ \bibinfo {author} {\bibfnamefont {E.}~\bibnamefont {Gorbatov}},\ }\href {\doibase 10.1088/1475-7516/2003/06/001} {\bibfield  {journal} {\bibinfo  {journal} {JCAP}\ }\textbf {\bibinfo {volume} {0306}},\ \bibinfo {pages} {001} (\bibinfo {year} {2003})},\ \Eprint {http://arxiv.org/abs/hep-th/0303252} {arXiv:hep-th/0303252 [hep-th]} \BibitemShut {NoStop}%
\bibitem [{\citenamefont {Arkani-Hamed}\ \emph {et~al.}(2007)\citenamefont {Arkani-Hamed}, \citenamefont {Motl}, \citenamefont {Nicolis},\ and\ \citenamefont {Vafa}}]{Arkani-Hamed:2006emk}%
  \BibitemOpen
  \bibfield  {author} {\bibinfo {author} {\bibfnamefont {N.}~\bibnamefont {Arkani-Hamed}}, \bibinfo {author} {\bibfnamefont {L.}~\bibnamefont {Motl}}, \bibinfo {author} {\bibfnamefont {A.}~\bibnamefont {Nicolis}}, \ and\ \bibinfo {author} {\bibfnamefont {C.}~\bibnamefont {Vafa}},\ }\href {\doibase 10.1088/1126-6708/2007/06/060} {\bibfield  {journal} {\bibinfo  {journal} {JHEP}\ }\textbf {\bibinfo {volume} {06}},\ \bibinfo {pages} {060} (\bibinfo {year} {2007})},\ \Eprint {http://arxiv.org/abs/hep-th/0601001} {arXiv:hep-th/0601001} \BibitemShut {NoStop}%
\bibitem [{\citenamefont {Rudelius}(2015)}]{Rudelius:2015xta}%
  \BibitemOpen
  \bibfield  {author} {\bibinfo {author} {\bibfnamefont {T.}~\bibnamefont {Rudelius}},\ }\href {\doibase 10.1088/1475-7516/2015/09/020, 10.1088/1475-7516/2015/9/020} {\bibfield  {journal} {\bibinfo  {journal} {JCAP}\ }\textbf {\bibinfo {volume} {1509}},\ \bibinfo {pages} {020} (\bibinfo {year} {2015})},\ \Eprint {http://arxiv.org/abs/1503.00795} {arXiv:1503.00795 [hep-th]} \BibitemShut {NoStop}%
\bibitem [{\citenamefont {Brown}\ \emph {et~al.}(2015)\citenamefont {Brown}, \citenamefont {Cottrell}, \citenamefont {Shiu},\ and\ \citenamefont {Soler}}]{Brown:2015iha}%
  \BibitemOpen
  \bibfield  {author} {\bibinfo {author} {\bibfnamefont {J.}~\bibnamefont {Brown}}, \bibinfo {author} {\bibfnamefont {W.}~\bibnamefont {Cottrell}}, \bibinfo {author} {\bibfnamefont {G.}~\bibnamefont {Shiu}}, \ and\ \bibinfo {author} {\bibfnamefont {P.}~\bibnamefont {Soler}},\ }\href {\doibase 10.1007/JHEP10(2015)023} {\bibfield  {journal} {\bibinfo  {journal} {JHEP}\ }\textbf {\bibinfo {volume} {10}},\ \bibinfo {pages} {023} (\bibinfo {year} {2015})},\ \Eprint {http://arxiv.org/abs/1503.04783} {arXiv:1503.04783 [hep-th]} \BibitemShut {NoStop}%
\bibitem [{\citenamefont {Hebecker}\ \emph {et~al.}(2016)\citenamefont {Hebecker}, \citenamefont {Rompineve},\ and\ \citenamefont {Westphal}}]{Hebecker:2015zss}%
  \BibitemOpen
  \bibfield  {author} {\bibinfo {author} {\bibfnamefont {A.}~\bibnamefont {Hebecker}}, \bibinfo {author} {\bibfnamefont {F.}~\bibnamefont {Rompineve}}, \ and\ \bibinfo {author} {\bibfnamefont {A.}~\bibnamefont {Westphal}},\ }\href {\doibase 10.1007/JHEP04(2016)157} {\bibfield  {journal} {\bibinfo  {journal} {JHEP}\ }\textbf {\bibinfo {volume} {04}},\ \bibinfo {pages} {157} (\bibinfo {year} {2016})},\ \Eprint {http://arxiv.org/abs/1512.03768} {arXiv:1512.03768 [hep-th]} \BibitemShut {NoStop}%
\bibitem [{\citenamefont {Svrcek}\ and\ \citenamefont {Witten}(2006)}]{Svrcek:2006yi}%
  \BibitemOpen
  \bibfield  {author} {\bibinfo {author} {\bibfnamefont {P.}~\bibnamefont {Svrcek}}\ and\ \bibinfo {author} {\bibfnamefont {E.}~\bibnamefont {Witten}},\ }\href {\doibase 10.1088/1126-6708/2006/06/051} {\bibfield  {journal} {\bibinfo  {journal} {JHEP}\ }\textbf {\bibinfo {volume} {06}},\ \bibinfo {pages} {051} (\bibinfo {year} {2006})},\ \Eprint {http://arxiv.org/abs/hep-th/0605206} {arXiv:hep-th/0605206} \BibitemShut {NoStop}%
\bibitem [{\citenamefont {Conlon}(2006)}]{Conlon:2006tq}%
  \BibitemOpen
  \bibfield  {author} {\bibinfo {author} {\bibfnamefont {J.~P.}\ \bibnamefont {Conlon}},\ }\href {\doibase 10.1088/1126-6708/2006/05/078} {\bibfield  {journal} {\bibinfo  {journal} {JHEP}\ }\textbf {\bibinfo {volume} {05}},\ \bibinfo {pages} {078} (\bibinfo {year} {2006})},\ \Eprint {http://arxiv.org/abs/hep-th/0602233} {arXiv:hep-th/0602233} \BibitemShut {NoStop}%
\bibitem [{\citenamefont {{Rudelius}}(2023)}]{Rudelius:2022gyu}%
  \BibitemOpen
  \bibfield  {author} {\bibinfo {author} {\bibfnamefont {T.}~\bibnamefont {{Rudelius}}},\ }\href {\doibase 10.1088/1475-7516/2023/01/014} {\bibfield  {journal} {\bibinfo  {journal} {JCAP}\ }\textbf {\bibinfo {volume} {1}},\ \bibinfo {eid} {014} (\bibinfo {year} {2023})},\ \Eprint {http://arxiv.org/abs/2203.05575} {arXiv:2203.05575 [hep-th]} \BibitemShut {NoStop}%
\bibitem [{\citenamefont {Ooguri}\ and\ \citenamefont {Vafa}(2007)}]{Ooguri:2006in}%
  \BibitemOpen
  \bibfield  {author} {\bibinfo {author} {\bibfnamefont {H.}~\bibnamefont {Ooguri}}\ and\ \bibinfo {author} {\bibfnamefont {C.}~\bibnamefont {Vafa}},\ }\href {\doibase 10.1016/j.nuclphysb.2006.10.033} {\bibfield  {journal} {\bibinfo  {journal} {Nucl. Phys. B}\ }\textbf {\bibinfo {volume} {766}},\ \bibinfo {pages} {21} (\bibinfo {year} {2007})},\ \Eprint {http://arxiv.org/abs/hep-th/0605264} {arXiv:hep-th/0605264} \BibitemShut {NoStop}%
\bibitem [{\citenamefont {Baume}\ and\ \citenamefont {Palti}(2016)}]{Baume:2016psm}%
  \BibitemOpen
  \bibfield  {author} {\bibinfo {author} {\bibfnamefont {F.}~\bibnamefont {Baume}}\ and\ \bibinfo {author} {\bibfnamefont {E.}~\bibnamefont {Palti}},\ }\href {\doibase 10.1007/JHEP08(2016)043} {\bibfield  {journal} {\bibinfo  {journal} {JHEP}\ }\textbf {\bibinfo {volume} {08}},\ \bibinfo {pages} {043} (\bibinfo {year} {2016})},\ \Eprint {http://arxiv.org/abs/1602.06517} {arXiv:1602.06517 [hep-th]} \BibitemShut {NoStop}%
\bibitem [{\citenamefont {Klaewer}\ and\ \citenamefont {Palti}(2017)}]{Klaewer:2016kiy}%
  \BibitemOpen
  \bibfield  {author} {\bibinfo {author} {\bibfnamefont {D.}~\bibnamefont {Klaewer}}\ and\ \bibinfo {author} {\bibfnamefont {E.}~\bibnamefont {Palti}},\ }\href {\doibase 10.1007/JHEP01(2017)088} {\bibfield  {journal} {\bibinfo  {journal} {JHEP}\ }\textbf {\bibinfo {volume} {01}},\ \bibinfo {pages} {088} (\bibinfo {year} {2017})},\ \Eprint {http://arxiv.org/abs/1610.00010} {arXiv:1610.00010 [hep-th]} \BibitemShut {NoStop}%
\bibitem [{\citenamefont {Blumenhagen}\ \emph {et~al.}(2017)\citenamefont {Blumenhagen}, \citenamefont {Valenzuela},\ and\ \citenamefont {Wolf}}]{Blumenhagen:2017cxt}%
  \BibitemOpen
  \bibfield  {author} {\bibinfo {author} {\bibfnamefont {R.}~\bibnamefont {Blumenhagen}}, \bibinfo {author} {\bibfnamefont {I.}~\bibnamefont {Valenzuela}}, \ and\ \bibinfo {author} {\bibfnamefont {F.}~\bibnamefont {Wolf}},\ }\href {\doibase 10.1007/JHEP07(2017)145} {\bibfield  {journal} {\bibinfo  {journal} {JHEP}\ }\textbf {\bibinfo {volume} {07}},\ \bibinfo {pages} {145} (\bibinfo {year} {2017})},\ \Eprint {http://arxiv.org/abs/1703.05776} {arXiv:1703.05776 [hep-th]} \BibitemShut {NoStop}%
\bibitem [{\citenamefont {Scalisi}\ and\ \citenamefont {Valenzuela}(2019)}]{Scalisi:2018eaz}%
  \BibitemOpen
  \bibfield  {author} {\bibinfo {author} {\bibfnamefont {M.}~\bibnamefont {Scalisi}}\ and\ \bibinfo {author} {\bibfnamefont {I.}~\bibnamefont {Valenzuela}},\ }\href {\doibase 10.1007/JHEP08(2019)160} {\bibfield  {journal} {\bibinfo  {journal} {JHEP}\ }\textbf {\bibinfo {volume} {08}},\ \bibinfo {pages} {160} (\bibinfo {year} {2019})},\ \Eprint {http://arxiv.org/abs/1812.07558} {arXiv:1812.07558 [hep-th]} \BibitemShut {NoStop}%
\bibitem [{\citenamefont {Broeckel}\ \emph {et~al.}(2021)\citenamefont {Broeckel}, \citenamefont {Cicoli}, \citenamefont {Maharana}, \citenamefont {Singh},\ and\ \citenamefont {Sinha}}]{Broeckel:2021dpz}%
  \BibitemOpen
  \bibfield  {author} {\bibinfo {author} {\bibfnamefont {I.}~\bibnamefont {Broeckel}}, \bibinfo {author} {\bibfnamefont {M.}~\bibnamefont {Cicoli}}, \bibinfo {author} {\bibfnamefont {A.}~\bibnamefont {Maharana}}, \bibinfo {author} {\bibfnamefont {K.}~\bibnamefont {Singh}}, \ and\ \bibinfo {author} {\bibfnamefont {K.}~\bibnamefont {Sinha}},\ }\href {\doibase 10.1007/JHEP01(2022)191} {\bibfield  {journal} {\bibinfo  {journal} {JHEP}\ }\textbf {\bibinfo {volume} {08}},\ \bibinfo {pages} {059} (\bibinfo {year} {2021})},\ \bibinfo {note} {[Addendum: JHEP 01, 191 (2022)]},\ \Eprint {http://arxiv.org/abs/2105.02889} {arXiv:2105.02889 [hep-th]} \BibitemShut {NoStop}%
\bibitem [{\citenamefont {Halverson}\ \emph {et~al.}(2019)\citenamefont {Halverson}, \citenamefont {Long}, \citenamefont {Nelson},\ and\ \citenamefont {Salinas}}]{Halverson:2019cmy}%
  \BibitemOpen
  \bibfield  {author} {\bibinfo {author} {\bibfnamefont {J.}~\bibnamefont {Halverson}}, \bibinfo {author} {\bibfnamefont {C.}~\bibnamefont {Long}}, \bibinfo {author} {\bibfnamefont {B.}~\bibnamefont {Nelson}}, \ and\ \bibinfo {author} {\bibfnamefont {G.}~\bibnamefont {Salinas}},\ }\href {\doibase 10.1103/PhysRevD.100.106010} {\bibfield  {journal} {\bibinfo  {journal} {Phys. Rev.}\ }\textbf {\bibinfo {volume} {D100}},\ \bibinfo {pages} {106010} (\bibinfo {year} {2019})},\ \Eprint {http://arxiv.org/abs/1909.05257} {arXiv:1909.05257 [hep-th]} \BibitemShut {NoStop}%
\bibitem [{\citenamefont {Mehta}\ \emph {et~al.}(2020)\citenamefont {Mehta}, \citenamefont {Demirtas}, \citenamefont {Long}, \citenamefont {Marsh}, \citenamefont {Mcallister},\ and\ \citenamefont {Stott}}]{Mehta:2020kwu}%
  \BibitemOpen
  \bibfield  {author} {\bibinfo {author} {\bibfnamefont {V.~M.}\ \bibnamefont {Mehta}}, \bibinfo {author} {\bibfnamefont {M.}~\bibnamefont {Demirtas}}, \bibinfo {author} {\bibfnamefont {C.}~\bibnamefont {Long}}, \bibinfo {author} {\bibfnamefont {D.~J.~E.}\ \bibnamefont {Marsh}}, \bibinfo {author} {\bibfnamefont {L.}~\bibnamefont {Mcallister}}, \ and\ \bibinfo {author} {\bibfnamefont {M.~J.}\ \bibnamefont {Stott}},\ }\href@noop {} {\  (\bibinfo {year} {2020})},\ \Eprint {http://arxiv.org/abs/2011.08693} {arXiv:2011.08693 [hep-th]} \BibitemShut {NoStop}%
\bibitem [{\citenamefont {Mehta}\ \emph {et~al.}(2021)\citenamefont {Mehta}, \citenamefont {Demirtas}, \citenamefont {Long}, \citenamefont {Marsh}, \citenamefont {McAllister},\ and\ \citenamefont {Stott}}]{Mehta:2021pwf}%
  \BibitemOpen
  \bibfield  {author} {\bibinfo {author} {\bibfnamefont {V.~M.}\ \bibnamefont {Mehta}}, \bibinfo {author} {\bibfnamefont {M.}~\bibnamefont {Demirtas}}, \bibinfo {author} {\bibfnamefont {C.}~\bibnamefont {Long}}, \bibinfo {author} {\bibfnamefont {D.~J.~E.}\ \bibnamefont {Marsh}}, \bibinfo {author} {\bibfnamefont {L.}~\bibnamefont {McAllister}}, \ and\ \bibinfo {author} {\bibfnamefont {M.~J.}\ \bibnamefont {Stott}},\ }\href {\doibase 10.1088/1475-7516/2021/07/033} {\bibfield  {journal} {\bibinfo  {journal} {JCAP}\ }\textbf {\bibinfo {volume} {07}},\ \bibinfo {pages} {033} (\bibinfo {year} {2021})},\ \Eprint {http://arxiv.org/abs/2103.06812} {arXiv:2103.06812 [hep-th]} \BibitemShut {NoStop}%
\bibitem [{\citenamefont {Demirtas}\ \emph {et~al.}(2023)\citenamefont {Demirtas}, \citenamefont {Gendler}, \citenamefont {Long}, \citenamefont {McAllister},\ and\ \citenamefont {Moritz}}]{Demirtas:2021gsq}%
  \BibitemOpen
  \bibfield  {author} {\bibinfo {author} {\bibfnamefont {M.}~\bibnamefont {Demirtas}}, \bibinfo {author} {\bibfnamefont {N.}~\bibnamefont {Gendler}}, \bibinfo {author} {\bibfnamefont {C.}~\bibnamefont {Long}}, \bibinfo {author} {\bibfnamefont {L.}~\bibnamefont {McAllister}}, \ and\ \bibinfo {author} {\bibfnamefont {J.}~\bibnamefont {Moritz}},\ }\href {\doibase 10.1007/JHEP06(2023)092} {\bibfield  {journal} {\bibinfo  {journal} {JHEP}\ }\textbf {\bibinfo {volume} {06}},\ \bibinfo {pages} {092} (\bibinfo {year} {2023})},\ \Eprint {http://arxiv.org/abs/2112.04503} {arXiv:2112.04503 [hep-th]} \BibitemShut {NoStop}%
\bibitem [{\citenamefont {Lesgourgues}(2011)}]{Lesgourgues:2011re}%
  \BibitemOpen
  \bibfield  {author} {\bibinfo {author} {\bibfnamefont {J.}~\bibnamefont {Lesgourgues}},\ }\href@noop {} {\  (\bibinfo {year} {2011})},\ \Eprint {http://arxiv.org/abs/1104.2932} {arXiv:1104.2932 [astro-ph.IM]} \BibitemShut {NoStop}%
\bibitem [{\citenamefont {Adame}\ \emph {et~al.}(2025{\natexlab{b}})\citenamefont {Adame} \emph {et~al.}}]{DESI:2024uvr}%
  \BibitemOpen
  \bibfield  {author} {\bibinfo {author} {\bibfnamefont {A.~G.}\ \bibnamefont {Adame}} \emph {et~al.} (\bibinfo {collaboration} {DESI}),\ }\href {\doibase 10.1088/1475-7516/2025/04/012} {\bibfield  {journal} {\bibinfo  {journal} {JCAP}\ }\textbf {\bibinfo {volume} {04}},\ \bibinfo {pages} {012} (\bibinfo {year} {2025}{\natexlab{b}})},\ \Eprint {http://arxiv.org/abs/2404.03000} {arXiv:2404.03000 [astro-ph.CO]} \BibitemShut {NoStop}%
\bibitem [{\citenamefont {Calderon}\ \emph {et~al.}(2024)\citenamefont {Calderon} \emph {et~al.}}]{DESI:2024aqx}%
  \BibitemOpen
  \bibfield  {author} {\bibinfo {author} {\bibfnamefont {R.}~\bibnamefont {Calderon}} \emph {et~al.} (\bibinfo {collaboration} {DESI}),\ }\href {\doibase 10.1088/1475-7516/2024/10/048} {\bibfield  {journal} {\bibinfo  {journal} {JCAP}\ }\textbf {\bibinfo {volume} {10}},\ \bibinfo {pages} {048} (\bibinfo {year} {2024})},\ \Eprint {http://arxiv.org/abs/2405.04216} {arXiv:2405.04216 [astro-ph.CO]} \BibitemShut {NoStop}%
\bibitem [{\citenamefont {Efstathiou}\ and\ \citenamefont {Gratton}(2021)}]{Efstathiou:2019mdh}%
  \BibitemOpen
  \bibfield  {author} {\bibinfo {author} {\bibfnamefont {G.}~\bibnamefont {Efstathiou}}\ and\ \bibinfo {author} {\bibfnamefont {S.}~\bibnamefont {Gratton}},\ }\href {\doibase 10.21105/astro.1910.00483} {\bibfield  {journal} {\bibinfo  {journal} {The Open Journal of Astrophysics}\ }\textbf {\bibinfo {volume} {4}} (\bibinfo {year} {2021}),\ 10.21105/astro.1910.00483},\ \Eprint {http://arxiv.org/abs/1910.00483} {arXiv:1910.00483 [astro-ph.CO]} \BibitemShut {NoStop}%
\bibitem [{\citenamefont {Rosenberg}\ \emph {et~al.}(2022)\citenamefont {Rosenberg}, \citenamefont {Gratton},\ and\ \citenamefont {Efstathiou}}]{Rosenberg:2022sdy}%
  \BibitemOpen
  \bibfield  {author} {\bibinfo {author} {\bibfnamefont {E.}~\bibnamefont {Rosenberg}}, \bibinfo {author} {\bibfnamefont {S.}~\bibnamefont {Gratton}}, \ and\ \bibinfo {author} {\bibfnamefont {G.}~\bibnamefont {Efstathiou}},\ }\href {\doibase 10.1093/mnras/stac2744} {\bibfield  {journal} {\bibinfo  {journal} {Mon. Not. Roy. Astron. Soc.}\ }\textbf {\bibinfo {volume} {517}},\ \bibinfo {pages} {4620} (\bibinfo {year} {2022})},\ \Eprint {http://arxiv.org/abs/2205.10869} {arXiv:2205.10869 [astro-ph.CO]} \BibitemShut {NoStop}%
\bibitem [{\citenamefont {Carron}\ \emph {et~al.}(2022)\citenamefont {Carron}, \citenamefont {Mirmelstein},\ and\ \citenamefont {Lewis}}]{Carron:2022eyg}%
  \BibitemOpen
  \bibfield  {author} {\bibinfo {author} {\bibfnamefont {J.}~\bibnamefont {Carron}}, \bibinfo {author} {\bibfnamefont {M.}~\bibnamefont {Mirmelstein}}, \ and\ \bibinfo {author} {\bibfnamefont {A.}~\bibnamefont {Lewis}},\ }\href {\doibase 10.1088/1475-7516/2022/09/039} {\bibfield  {journal} {\bibinfo  {journal} {JCAP}\ }\textbf {\bibinfo {volume} {09}},\ \bibinfo {pages} {039} (\bibinfo {year} {2022})},\ \Eprint {http://arxiv.org/abs/2206.07773} {arXiv:2206.07773 [astro-ph.CO]} \BibitemShut {NoStop}%
\bibitem [{\citenamefont {Scolnic}\ \emph {et~al.}(2022)\citenamefont {Scolnic} \emph {et~al.}}]{Scolnic:2021amr}%
  \BibitemOpen
  \bibfield  {author} {\bibinfo {author} {\bibfnamefont {D.}~\bibnamefont {Scolnic}} \emph {et~al.},\ }\href {\doibase 10.3847/1538-4357/ac8b7a} {\bibfield  {journal} {\bibinfo  {journal} {Astrophys. J.}\ }\textbf {\bibinfo {volume} {938}},\ \bibinfo {eid} {113} (\bibinfo {year} {2022})},\ \Eprint {http://arxiv.org/abs/2112.03863} {arXiv:2112.03863 [astro-ph.CO]} \BibitemShut {NoStop}%
\bibitem [{\citenamefont {Brout}\ \emph {et~al.}(2022)\citenamefont {Brout} \emph {et~al.}}]{Brout:2022vxf}%
  \BibitemOpen
  \bibfield  {author} {\bibinfo {author} {\bibfnamefont {D.}~\bibnamefont {Brout}} \emph {et~al.},\ }\href {\doibase 10.3847/1538-4357/ac8e04} {\bibfield  {journal} {\bibinfo  {journal} {Astrophys. J.}\ }\textbf {\bibinfo {volume} {938}},\ \bibinfo {pages} {110} (\bibinfo {year} {2022})},\ \Eprint {http://arxiv.org/abs/2202.04077} {arXiv:2202.04077 [astro-ph.CO]} \BibitemShut {NoStop}%
\bibitem [{\citenamefont {Rubin}\ \emph {et~al.}(2025)\citenamefont {Rubin} \emph {et~al.}}]{Rubin:2023jdq}%
  \BibitemOpen
  \bibfield  {author} {\bibinfo {author} {\bibfnamefont {D.}~\bibnamefont {Rubin}} \emph {et~al.},\ }\href {\doibase 10.3847/1538-4357/adc0a5} {\bibfield  {journal} {\bibinfo  {journal} {Astrophys. J.}\ }\textbf {\bibinfo {volume} {986}},\ \bibinfo {eid} {231} (\bibinfo {year} {2025})},\ \Eprint {http://arxiv.org/abs/2311.12098} {arXiv:2311.12098 [astro-ph.CO]} \BibitemShut {NoStop}%
\bibitem [{\citenamefont {Abbott}\ \emph {et~al.}(2024)\citenamefont {Abbott} \emph {et~al.}}]{DES:2024jxu}%
  \BibitemOpen
  \bibfield  {author} {\bibinfo {author} {\bibfnamefont {T.~M.~C.}\ \bibnamefont {Abbott}} \emph {et~al.} (\bibinfo {collaboration} {DES}),\ }\href {\doibase 10.3847/2041-8213/ad6f9f} {\bibfield  {journal} {\bibinfo  {journal} {Astrophys. J. Lett.}\ }\textbf {\bibinfo {volume} {973}},\ \bibinfo {pages} {L14} (\bibinfo {year} {2024})},\ \Eprint {http://arxiv.org/abs/2401.02929} {arXiv:2401.02929 [astro-ph.CO]} \BibitemShut {NoStop}%
\bibitem [{\citenamefont {{Popovic}}\ \emph {et~al.}(2025{\natexlab{a}})\citenamefont {{Popovic}} \emph {et~al.}}]{Popovic2025:Dovekie_1}%
  \BibitemOpen
  \bibfield  {author} {\bibinfo {author} {\bibfnamefont {B.}~\bibnamefont {{Popovic}}} \emph {et~al.},\ }\href@noop {} {\  (\bibinfo {year} {2025}{\natexlab{a}})},\ \Eprint {http://arxiv.org/abs/2506.05471} {arXiv:2506.05471 [astro-ph.CO]} \BibitemShut {NoStop}%
\bibitem [{\citenamefont {{Popovic}}\ \emph {et~al.}(2025{\natexlab{b}})\citenamefont {{Popovic}} \emph {et~al.}}]{Popovic2025:Dovekie_2}%
  \BibitemOpen
  \bibfield  {author} {\bibinfo {author} {\bibfnamefont {B.}~\bibnamefont {{Popovic}}} \emph {et~al.},\ }\href@noop {} {\  (\bibinfo {year} {2025}{\natexlab{b}})},\ \Eprint {http://arxiv.org/abs/2511.07517} {arXiv:2511.07517 [astro-ph.CO]} \BibitemShut {NoStop}%
\bibitem [{\citenamefont {Torrado}\ and\ \citenamefont {Lewis}(2021)}]{Torrado:2020dgo}%
  \BibitemOpen
  \bibfield  {author} {\bibinfo {author} {\bibfnamefont {J.}~\bibnamefont {Torrado}}\ and\ \bibinfo {author} {\bibfnamefont {A.}~\bibnamefont {Lewis}},\ }\href {\doibase 10.1088/1475-7516/2021/05/057} {\bibfield  {journal} {\bibinfo  {journal} {JCAP}\ }\textbf {\bibinfo {volume} {05}},\ \bibinfo {pages} {057} (\bibinfo {year} {2021})},\ \Eprint {http://arxiv.org/abs/2005.05290} {arXiv:2005.05290 [astro-ph.IM]} \BibitemShut {NoStop}%
\bibitem [{\citenamefont {Blas}\ \emph {et~al.}(2011)\citenamefont {Blas}, \citenamefont {Lesgourgues},\ and\ \citenamefont {Tram}}]{Blas:2011rf}%
  \BibitemOpen
  \bibfield  {author} {\bibinfo {author} {\bibfnamefont {D.}~\bibnamefont {Blas}}, \bibinfo {author} {\bibfnamefont {J.}~\bibnamefont {Lesgourgues}}, \ and\ \bibinfo {author} {\bibfnamefont {T.}~\bibnamefont {Tram}},\ }\href {\doibase 10.1088/1475-7516/2011/07/034} {\bibfield  {journal} {\bibinfo  {journal} {JCAP}\ }\textbf {\bibinfo {volume} {1107}},\ \bibinfo {pages} {034} (\bibinfo {year} {2011})},\ \Eprint {http://arxiv.org/abs/1104.2933} {arXiv:1104.2933} \BibitemShut {NoStop}%
\bibitem [{\citenamefont {{Neal}}(2005)}]{2005math......2099N}%
  \BibitemOpen
  \bibfield  {author} {\bibinfo {author} {\bibfnamefont {R.~M.}\ \bibnamefont {{Neal}}},\ }\href@noop {} {\  (\bibinfo {year} {2005})},\ \Eprint {http://arxiv.org/abs/math/0502099} {arXiv:math/0502099 [math.ST]} \BibitemShut {NoStop}%
\bibitem [{\citenamefont {Gelman}\ and\ \citenamefont {Rubin}(1992)}]{10.1214/ss/1177011136}%
  \BibitemOpen
  \bibfield  {author} {\bibinfo {author} {\bibfnamefont {A.}~\bibnamefont {Gelman}}\ and\ \bibinfo {author} {\bibfnamefont {D.~B.}\ \bibnamefont {Rubin}},\ }\href {\doibase 10.1214/ss/1177011136} {\bibfield  {journal} {\bibinfo  {journal} {Statistical Science}\ }\textbf {\bibinfo {volume} {7}},\ \bibinfo {pages} {457 } (\bibinfo {year} {1992})}\BibitemShut {NoStop}%
\bibitem [{\citenamefont {Lewis}(2025)}]{Lewis:2019xzd}%
  \BibitemOpen
  \bibfield  {author} {\bibinfo {author} {\bibfnamefont {A.}~\bibnamefont {Lewis}},\ }\href {\doibase 10.1088/1475-7516/2025/08/025} {\bibfield  {journal} {\bibinfo  {journal} {JCAP}\ }\textbf {\bibinfo {volume} {8}},\ \bibinfo {eid} {025} (\bibinfo {year} {2025})},\ \Eprint {http://arxiv.org/abs/1910.13970} {arXiv:1910.13970 [astro-ph.IM]} \BibitemShut {NoStop}%
\bibitem [{\citenamefont {Ivanov}\ \emph {et~al.}(2020)\citenamefont {Ivanov}, \citenamefont {Simonovi\'c},\ and\ \citenamefont {Zaldarriaga}}]{Ivanov:2019pdj}%
  \BibitemOpen
  \bibfield  {author} {\bibinfo {author} {\bibfnamefont {M.~M.}\ \bibnamefont {Ivanov}}, \bibinfo {author} {\bibfnamefont {M.}~\bibnamefont {Simonovi\'c}}, \ and\ \bibinfo {author} {\bibfnamefont {M.}~\bibnamefont {Zaldarriaga}},\ }\href {\doibase 10.1088/1475-7516/2020/05/042} {\bibfield  {journal} {\bibinfo  {journal} {JCAP}\ }\textbf {\bibinfo {volume} {05}},\ \bibinfo {pages} {042} (\bibinfo {year} {2020})},\ \Eprint {http://arxiv.org/abs/1909.05277} {arXiv:1909.05277 [astro-ph.CO]} \BibitemShut {NoStop}%
\bibitem [{\citenamefont {Chudaykin}\ \emph {et~al.}(2020)\citenamefont {Chudaykin}, \citenamefont {Ivanov},\ and\ \citenamefont {Simonovi\'c}}]{Chudaykin:2020aoj}%
  \BibitemOpen
  \bibfield  {author} {\bibinfo {author} {\bibfnamefont {A.}~\bibnamefont {Chudaykin}}, \bibinfo {author} {\bibfnamefont {M.~M.}\ \bibnamefont {Ivanov}}, \ and\ \bibinfo {author} {\bibfnamefont {M.}~\bibnamefont {Simonovi\'c}},\ }\href {\doibase 10.1103/PhysRevD.102.063533} {\bibfield  {journal} {\bibinfo  {journal} {Phys. Rev. D}\ }\textbf {\bibinfo {volume} {102}},\ \bibinfo {eid} {063533} (\bibinfo {year} {2020})},\ \Eprint {http://arxiv.org/abs/2004.10607} {arXiv:2004.10607 [astro-ph.CO]} \BibitemShut {NoStop}%
\bibitem [{\citenamefont {Chudaykin}\ \emph {et~al.}(2021)\citenamefont {Chudaykin}, \citenamefont {Dolgikh},\ and\ \citenamefont {Ivanov}}]{Chudaykin:2020ghx}%
  \BibitemOpen
  \bibfield  {author} {\bibinfo {author} {\bibfnamefont {A.}~\bibnamefont {Chudaykin}}, \bibinfo {author} {\bibfnamefont {K.}~\bibnamefont {Dolgikh}}, \ and\ \bibinfo {author} {\bibfnamefont {M.~M.}\ \bibnamefont {Ivanov}},\ }\href {\doibase 10.1103/PhysRevD.103.023507} {\bibfield  {journal} {\bibinfo  {journal} {Phys. Rev. D}\ }\textbf {\bibinfo {volume} {103}},\ \bibinfo {pages} {023507} (\bibinfo {year} {2021})},\ \Eprint {http://arxiv.org/abs/2009.10106} {arXiv:2009.10106 [astro-ph.CO]} \BibitemShut {NoStop}%
\bibitem [{\citenamefont {Ivanov}\ \emph {et~al.}(2022)\citenamefont {Ivanov}, \citenamefont {Philcox}, \citenamefont {Nishimichi}, \citenamefont {Simonovi\'c}, \citenamefont {Takada},\ and\ \citenamefont {Zaldarriaga}}]{Ivanov:2021kcd}%
  \BibitemOpen
  \bibfield  {author} {\bibinfo {author} {\bibfnamefont {M.~M.}\ \bibnamefont {Ivanov}}, \bibinfo {author} {\bibfnamefont {O.~H.~E.}\ \bibnamefont {Philcox}}, \bibinfo {author} {\bibfnamefont {T.}~\bibnamefont {Nishimichi}}, \bibinfo {author} {\bibfnamefont {M.}~\bibnamefont {Simonovi\'c}}, \bibinfo {author} {\bibfnamefont {M.}~\bibnamefont {Takada}}, \ and\ \bibinfo {author} {\bibfnamefont {M.}~\bibnamefont {Zaldarriaga}},\ }\href {\doibase 10.1103/PhysRevD.105.063512} {\bibfield  {journal} {\bibinfo  {journal} {Phys. Rev. D}\ }\textbf {\bibinfo {volume} {105}},\ \bibinfo {pages} {063512} (\bibinfo {year} {2022})},\ \Eprint {http://arxiv.org/abs/2110.10161} {arXiv:2110.10161 [astro-ph.CO]} \BibitemShut {NoStop}%
\bibitem [{\citenamefont {Ivanov}(2021)}]{Ivanov:2021zmi}%
  \BibitemOpen
  \bibfield  {author} {\bibinfo {author} {\bibfnamefont {M.~M.}\ \bibnamefont {Ivanov}},\ }\href {\doibase 10.1103/PhysRevD.104.103514} {\bibfield  {journal} {\bibinfo  {journal} {Phys. Rev. D}\ }\textbf {\bibinfo {volume} {104}},\ \bibinfo {pages} {103514} (\bibinfo {year} {2021})},\ \Eprint {http://arxiv.org/abs/2106.12580} {arXiv:2106.12580 [astro-ph.CO]} \BibitemShut {NoStop}%
\bibitem [{\citenamefont {Philcox}\ and\ \citenamefont {Ivanov}(2022)}]{Philcox:2021kcw}%
  \BibitemOpen
  \bibfield  {author} {\bibinfo {author} {\bibfnamefont {O.~H.~E.}\ \bibnamefont {Philcox}}\ and\ \bibinfo {author} {\bibfnamefont {M.~M.}\ \bibnamefont {Ivanov}},\ }\href {\doibase 10.1103/PhysRevD.105.043517} {\bibfield  {journal} {\bibinfo  {journal} {Phys. Rev. D}\ }\textbf {\bibinfo {volume} {105}},\ \bibinfo {pages} {043517} (\bibinfo {year} {2022})},\ \Eprint {http://arxiv.org/abs/2112.04515} {arXiv:2112.04515 [astro-ph.CO]} \BibitemShut {NoStop}%
\bibitem [{\citenamefont {Chudaykin}\ and\ \citenamefont {Ivanov}(2023)}]{Chudaykin:2022nru}%
  \BibitemOpen
  \bibfield  {author} {\bibinfo {author} {\bibfnamefont {A.}~\bibnamefont {Chudaykin}}\ and\ \bibinfo {author} {\bibfnamefont {M.~M.}\ \bibnamefont {Ivanov}},\ }\href {\doibase 10.1103/PhysRevD.107.043518} {\bibfield  {journal} {\bibinfo  {journal} {Phys. Rev. D}\ }\textbf {\bibinfo {volume} {107}},\ \bibinfo {pages} {043518} (\bibinfo {year} {2023})},\ \Eprint {http://arxiv.org/abs/2210.17044} {arXiv:2210.17044 [astro-ph.CO]} \BibitemShut {NoStop}%
\bibitem [{\citenamefont {D'Amico}\ \emph {et~al.}(2020)\citenamefont {D'Amico}, \citenamefont {Gleyzes}, \citenamefont {Kokron}, \citenamefont {Markovic}, \citenamefont {Senatore}, \citenamefont {Zhang}, \citenamefont {Beutler},\ and\ \citenamefont {Gil-Marín}}]{DAmico:2019fhj}%
  \BibitemOpen
  \bibfield  {author} {\bibinfo {author} {\bibfnamefont {G.}~\bibnamefont {D'Amico}}, \bibinfo {author} {\bibfnamefont {J.}~\bibnamefont {Gleyzes}}, \bibinfo {author} {\bibfnamefont {N.}~\bibnamefont {Kokron}}, \bibinfo {author} {\bibfnamefont {D.}~\bibnamefont {Markovic}}, \bibinfo {author} {\bibfnamefont {L.}~\bibnamefont {Senatore}}, \bibinfo {author} {\bibfnamefont {P.}~\bibnamefont {Zhang}}, \bibinfo {author} {\bibfnamefont {F.}~\bibnamefont {Beutler}}, \ and\ \bibinfo {author} {\bibfnamefont {H.}~\bibnamefont {Gil-Marín}},\ }\href {\doibase 10.1088/1475-7516/2020/05/005} {\bibfield  {journal} {\bibinfo  {journal} {JCAP}\ }\textbf {\bibinfo {volume} {5}},\ \bibinfo {eid} {005} (\bibinfo {year} {2020})},\ \Eprint {http://arxiv.org/abs/1909.05271} {arXiv:1909.05271 [astro-ph.CO]} \BibitemShut {NoStop}%
\bibitem [{\citenamefont {D'Amico}\ \emph {et~al.}(2021)\citenamefont {D'Amico}, \citenamefont {Senatore},\ and\ \citenamefont {Zhang}}]{DAmico:2020kxu}%
  \BibitemOpen
  \bibfield  {author} {\bibinfo {author} {\bibfnamefont {G.}~\bibnamefont {D'Amico}}, \bibinfo {author} {\bibfnamefont {L.}~\bibnamefont {Senatore}}, \ and\ \bibinfo {author} {\bibfnamefont {P.}~\bibnamefont {Zhang}},\ }\href {\doibase 10.1088/1475-7516/2021/01/006} {\bibfield  {journal} {\bibinfo  {journal} {JCAP}\ }\textbf {\bibinfo {volume} {01}},\ \bibinfo {pages} {006} (\bibinfo {year} {2021})},\ \Eprint {http://arxiv.org/abs/2003.07956} {arXiv:2003.07956 [astro-ph.CO]} \BibitemShut {NoStop}%
\bibitem [{\citenamefont {Chen}\ \emph {et~al.}(2022)\citenamefont {Chen}, \citenamefont {Vlah},\ and\ \citenamefont {White}}]{Chen:2021wdi}%
  \BibitemOpen
  \bibfield  {author} {\bibinfo {author} {\bibfnamefont {S.-F.}\ \bibnamefont {Chen}}, \bibinfo {author} {\bibfnamefont {Z.}~\bibnamefont {Vlah}}, \ and\ \bibinfo {author} {\bibfnamefont {M.}~\bibnamefont {White}},\ }\href {\doibase 10.1088/1475-7516/2022/02/008} {\bibfield  {journal} {\bibinfo  {journal} {JCAP}\ }\textbf {\bibinfo {volume} {02}},\ \bibinfo {pages} {008} (\bibinfo {year} {2022})},\ \Eprint {http://arxiv.org/abs/2110.05530} {arXiv:2110.05530 [astro-ph.CO]} \BibitemShut {NoStop}%
\bibitem [{\citenamefont {Zhang}\ \emph {et~al.}(2022)\citenamefont {Zhang}, \citenamefont {D'Amico}, \citenamefont {Senatore}, \citenamefont {Zhao},\ and\ \citenamefont {Cai}}]{Zhang:2021yna}%
  \BibitemOpen
  \bibfield  {author} {\bibinfo {author} {\bibfnamefont {P.}~\bibnamefont {Zhang}}, \bibinfo {author} {\bibfnamefont {G.}~\bibnamefont {D'Amico}}, \bibinfo {author} {\bibfnamefont {L.}~\bibnamefont {Senatore}}, \bibinfo {author} {\bibfnamefont {C.}~\bibnamefont {Zhao}}, \ and\ \bibinfo {author} {\bibfnamefont {Y.}~\bibnamefont {Cai}},\ }\href {\doibase 10.1088/1475-7516/2022/02/036} {\bibfield  {journal} {\bibinfo  {journal} {JCAP}\ }\textbf {\bibinfo {volume} {02}},\ \bibinfo {pages} {036} (\bibinfo {year} {2022})},\ \Eprint {http://arxiv.org/abs/2110.07539} {arXiv:2110.07539 [astro-ph.CO]} \BibitemShut {NoStop}%
\bibitem [{\citenamefont {Chudaykin}\ \emph {et~al.}(2024)\citenamefont {Chudaykin}, \citenamefont {Ivanov},\ and\ \citenamefont {Nishimichi}}]{Chudaykin:2024wlw}%
  \BibitemOpen
  \bibfield  {author} {\bibinfo {author} {\bibfnamefont {A.}~\bibnamefont {Chudaykin}}, \bibinfo {author} {\bibfnamefont {M.~M.}\ \bibnamefont {Ivanov}}, \ and\ \bibinfo {author} {\bibfnamefont {T.}~\bibnamefont {Nishimichi}},\ }\href@noop {} {\  (\bibinfo {year} {2024})},\ \Eprint {http://arxiv.org/abs/2410.16358} {arXiv:2410.16358 [astro-ph.CO]} \BibitemShut {NoStop}%
\bibitem [{\citenamefont {Ivanov}\ \emph {et~al.}(2024{\natexlab{c}})\citenamefont {Ivanov}, \citenamefont {Toomey},\ and\ \citenamefont {Kara\c{c}ayl\i{}}}]{Ivanov:2024jtl}%
  \BibitemOpen
  \bibfield  {author} {\bibinfo {author} {\bibfnamefont {M.~M.}\ \bibnamefont {Ivanov}}, \bibinfo {author} {\bibfnamefont {M.~W.}\ \bibnamefont {Toomey}}, \ and\ \bibinfo {author} {\bibfnamefont {N.~G.}\ \bibnamefont {Kara\c{c}ayl\i{}}},\ }\href {\doibase 10.1103/PhysRevLett.134.091001} {\bibfield  {journal} {\bibinfo  {journal} {Phys. Rev. Lett.}\ }\textbf {\bibinfo {volume} {134}},\ \bibinfo {eid} {091001} (\bibinfo {year} {2024}{\natexlab{c}})},\ \Eprint {http://arxiv.org/abs/2405.13208} {arXiv:2405.13208 [astro-ph.CO]} \BibitemShut {NoStop}%
\bibitem [{\citenamefont {Ivanov}(2025)}]{Ivanov:2025qie}%
  \BibitemOpen
  \bibfield  {author} {\bibinfo {author} {\bibfnamefont {M.~M.}\ \bibnamefont {Ivanov}},\ }\href@noop {} {\  (\bibinfo {year} {2025})},\ \Eprint {http://arxiv.org/abs/2503.07270} {arXiv:2503.07270 [astro-ph.CO]} \BibitemShut {NoStop}%
\bibitem [{\citenamefont {Sullivan}\ \emph {et~al.}(2021)\citenamefont {Sullivan}, \citenamefont {Seljak},\ and\ \citenamefont {Singh}}]{Sullivan:2021sof}%
  \BibitemOpen
  \bibfield  {author} {\bibinfo {author} {\bibfnamefont {J.~M.}\ \bibnamefont {Sullivan}}, \bibinfo {author} {\bibfnamefont {U.}~\bibnamefont {Seljak}}, \ and\ \bibinfo {author} {\bibfnamefont {S.}~\bibnamefont {Singh}},\ }\href {\doibase 10.1088/1475-7516/2021/11/026} {\bibfield  {journal} {\bibinfo  {journal} {JCAP}\ }\textbf {\bibinfo {volume} {11}},\ \bibinfo {pages} {026} (\bibinfo {year} {2021})},\ \Eprint {http://arxiv.org/abs/2104.10676} {arXiv:2104.10676 [astro-ph.CO]} \BibitemShut {NoStop}%
\bibitem [{\citenamefont {Akitsu}(2024)}]{Akitsu:2024lyt}%
  \BibitemOpen
  \bibfield  {author} {\bibinfo {author} {\bibfnamefont {K.}~\bibnamefont {Akitsu}},\ }\href@noop {} {\  (\bibinfo {year} {2024})},\ \Eprint {http://arxiv.org/abs/2410.08998} {arXiv:2410.08998 [astro-ph.CO]} \BibitemShut {NoStop}%
\bibitem [{\citenamefont {Cabass}\ \emph {et~al.}(2025)\citenamefont {Cabass}, \citenamefont {Philcox}, \citenamefont {Ivanov}, \citenamefont {Akitsu}, \citenamefont {Chen}, \citenamefont {Simonovi{\'c}},\ and\ \citenamefont {Zaldarriaga}}]{Cabass:2024wob}%
  \BibitemOpen
  \bibfield  {author} {\bibinfo {author} {\bibfnamefont {G.}~\bibnamefont {Cabass}}, \bibinfo {author} {\bibfnamefont {O.~H.~E.}\ \bibnamefont {Philcox}}, \bibinfo {author} {\bibfnamefont {M.~M.}\ \bibnamefont {Ivanov}}, \bibinfo {author} {\bibfnamefont {K.}~\bibnamefont {Akitsu}}, \bibinfo {author} {\bibfnamefont {S.-F.}\ \bibnamefont {Chen}}, \bibinfo {author} {\bibfnamefont {M.}~\bibnamefont {Simonovi{\'c}}}, \ and\ \bibinfo {author} {\bibfnamefont {M.}~\bibnamefont {Zaldarriaga}},\ }\href {\doibase 10.1103/PhysRevD.111.063510} {\bibfield  {journal} {\bibinfo  {journal} {Phys. Rev. D}\ }\textbf {\bibinfo {volume} {111}},\ \bibinfo {pages} {063510} (\bibinfo {year} {2025})},\ \Eprint {http://arxiv.org/abs/2404.01894} {arXiv:2404.01894 [astro-ph.CO]} \BibitemShut {NoStop}%
\bibitem [{\citenamefont {Zhang}\ \emph {et~al.}(2025{\natexlab{a}})\citenamefont {Zhang}, \citenamefont {Bonici}, \citenamefont {D'Amico}, \citenamefont {Paradiso},\ and\ \citenamefont {Percival}}]{Zhang:2024thl}%
  \BibitemOpen
  \bibfield  {author} {\bibinfo {author} {\bibfnamefont {H.}~\bibnamefont {Zhang}}, \bibinfo {author} {\bibfnamefont {M.}~\bibnamefont {Bonici}}, \bibinfo {author} {\bibfnamefont {G.}~\bibnamefont {D'Amico}}, \bibinfo {author} {\bibfnamefont {S.}~\bibnamefont {Paradiso}}, \ and\ \bibinfo {author} {\bibfnamefont {W.~J.}\ \bibnamefont {Percival}},\ }\href {\doibase 10.1088/1475-7516/2025/04/041} {\bibfield  {journal} {\bibinfo  {journal} {JCAP}\ }\textbf {\bibinfo {volume} {04}},\ \bibinfo {pages} {041} (\bibinfo {year} {2025}{\natexlab{a}})},\ \Eprint {http://arxiv.org/abs/2409.12937} {arXiv:2409.12937 [astro-ph.CO]} \BibitemShut {NoStop}%
\bibitem [{\citenamefont {Zhang}\ \emph {et~al.}(2025{\natexlab{b}})\citenamefont {Zhang} \emph {et~al.}}]{DESI:2025wzd}%
  \BibitemOpen
  \bibfield  {author} {\bibinfo {author} {\bibfnamefont {H.}~\bibnamefont {Zhang}} \emph {et~al.} (\bibinfo {collaboration} {DESI}),\ }\href {\doibase 10.1088/1475-7516/2025/11/049} {\bibfield  {journal} {\bibinfo  {journal} {JCAP}\ }\textbf {\bibinfo {volume} {11}},\ \bibinfo {pages} {049} (\bibinfo {year} {2025}{\natexlab{b}})},\ \Eprint {http://arxiv.org/abs/2504.10407} {arXiv:2504.10407 [astro-ph.CO]} \BibitemShut {NoStop}%
\bibitem [{\citenamefont {Chudaykin}\ \emph {et~al.}()\citenamefont {Chudaykin}, \citenamefont {Ivanov},\ and\ \citenamefont {Philcox}}]{Chudaykin:2025auz}%
  \BibitemOpen
  \bibfield  {author} {\bibinfo {author} {\bibfnamefont {A.}~\bibnamefont {Chudaykin}}, \bibinfo {author} {\bibfnamefont {M.~M.}\ \bibnamefont {Ivanov}}, \ and\ \bibinfo {author} {\bibfnamefont {O.~H.~E.}\ \bibnamefont {Philcox}},\ }\href@noop {} {\ }\Eprint {http://arxiv.org/abs/{to appear}} {{to appear}} \BibitemShut {NoStop}%
\bibitem [{\citenamefont {Goldstein}\ \emph {et~al.}(2023)\citenamefont {Goldstein}, \citenamefont {Hill}, \citenamefont {Ir\v{s}i\v{c}},\ and\ \citenamefont {Sherwin}}]{Goldstein:2023gnw}%
  \BibitemOpen
  \bibfield  {author} {\bibinfo {author} {\bibfnamefont {S.}~\bibnamefont {Goldstein}}, \bibinfo {author} {\bibfnamefont {J.~C.}\ \bibnamefont {Hill}}, \bibinfo {author} {\bibfnamefont {V.}~\bibnamefont {Ir\v{s}i\v{c}}}, \ and\ \bibinfo {author} {\bibfnamefont {B.~D.}\ \bibnamefont {Sherwin}},\ }\href@noop {} {\  (\bibinfo {year} {2023})},\ \Eprint {http://arxiv.org/abs/2303.00746} {arXiv:2303.00746 [astro-ph.CO]} \BibitemShut {NoStop}%
\bibitem [{\citenamefont {Besuner}\ \emph {et~al.}(2025)\citenamefont {Besuner} \emph {et~al.}}]{Spec-S5:2025uom}%
  \BibitemOpen
  \bibfield  {author} {\bibinfo {author} {\bibfnamefont {R.}~\bibnamefont {Besuner}} \emph {et~al.} (\bibinfo {collaboration} {Spec-S5}),\ }\href@noop {} {\  (\bibinfo {year} {2025})},\ \Eprint {http://arxiv.org/abs/2503.07923} {arXiv:2503.07923 [astro-ph.CO]} \BibitemShut {NoStop}%
\end{thebibliography}%

\newpage 

\pagebreak
\widetext

\appendix

\newpage

\section{Extra Plots and Tables}\label{appendix}

\begin{figure}[!h]
    \centering
    \includegraphics[width=\linewidth]{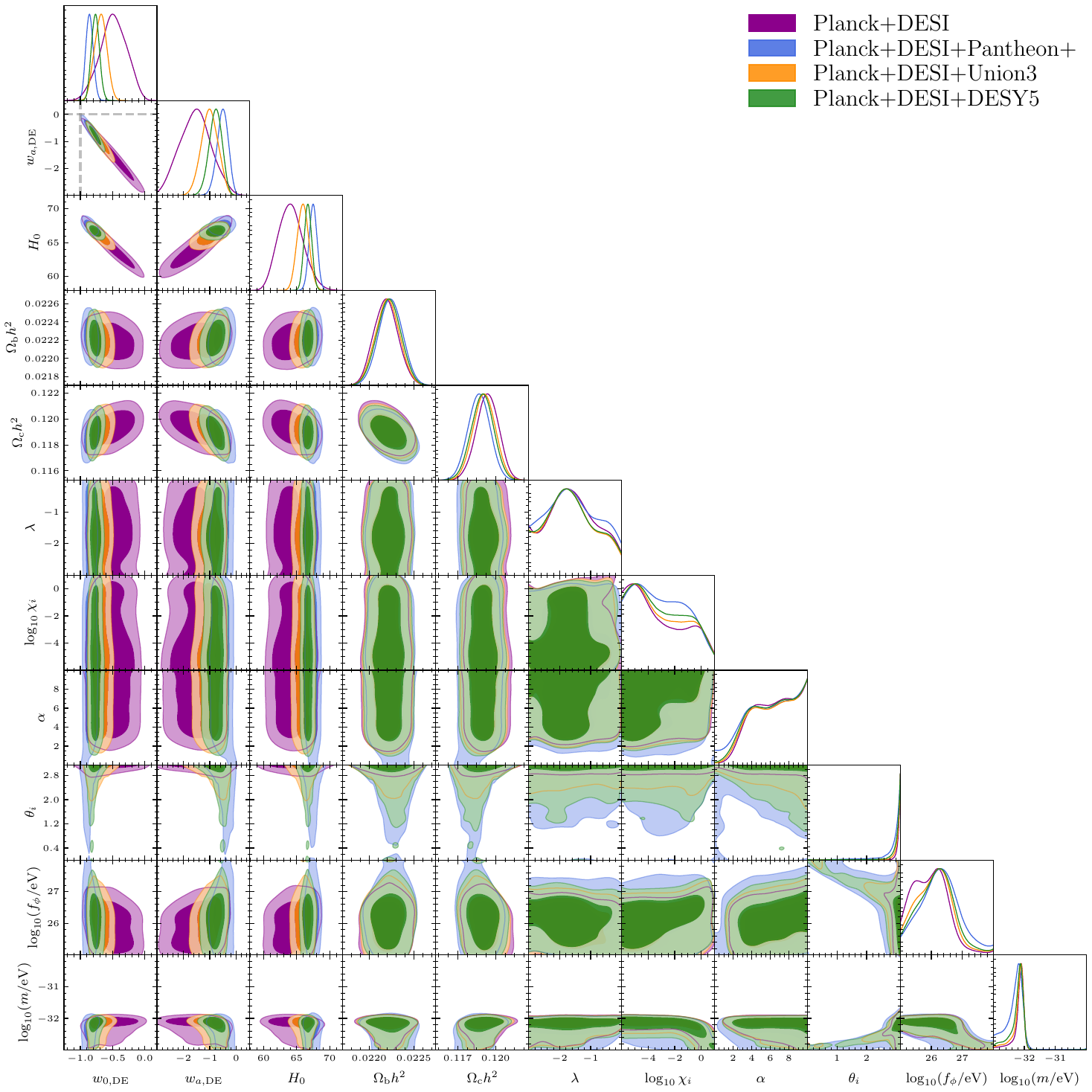}
    \caption{Full constraints at the 68\% and 95\% confidence levels for the kinetically mixed axion--dilaton (KMIX) model. The analysis combines \textit{Planck} PR4 CMB anisotropies (with lensing) and DESI DR2 BAO, shown in \textcolor{tabpurple}{purple}, and separately includes Type~Ia supernova datasets from Pantheon+ (\textcolor{tabblue}{blue}), Union~3 (\textcolor{taborange}{orange}), and DES~Y5 (\textcolor{tabgreen}{green}).}
    \label{fig:placeholder_2}
\end{figure}

\begin{figure}
    \centering
    \includegraphics[width=\linewidth]{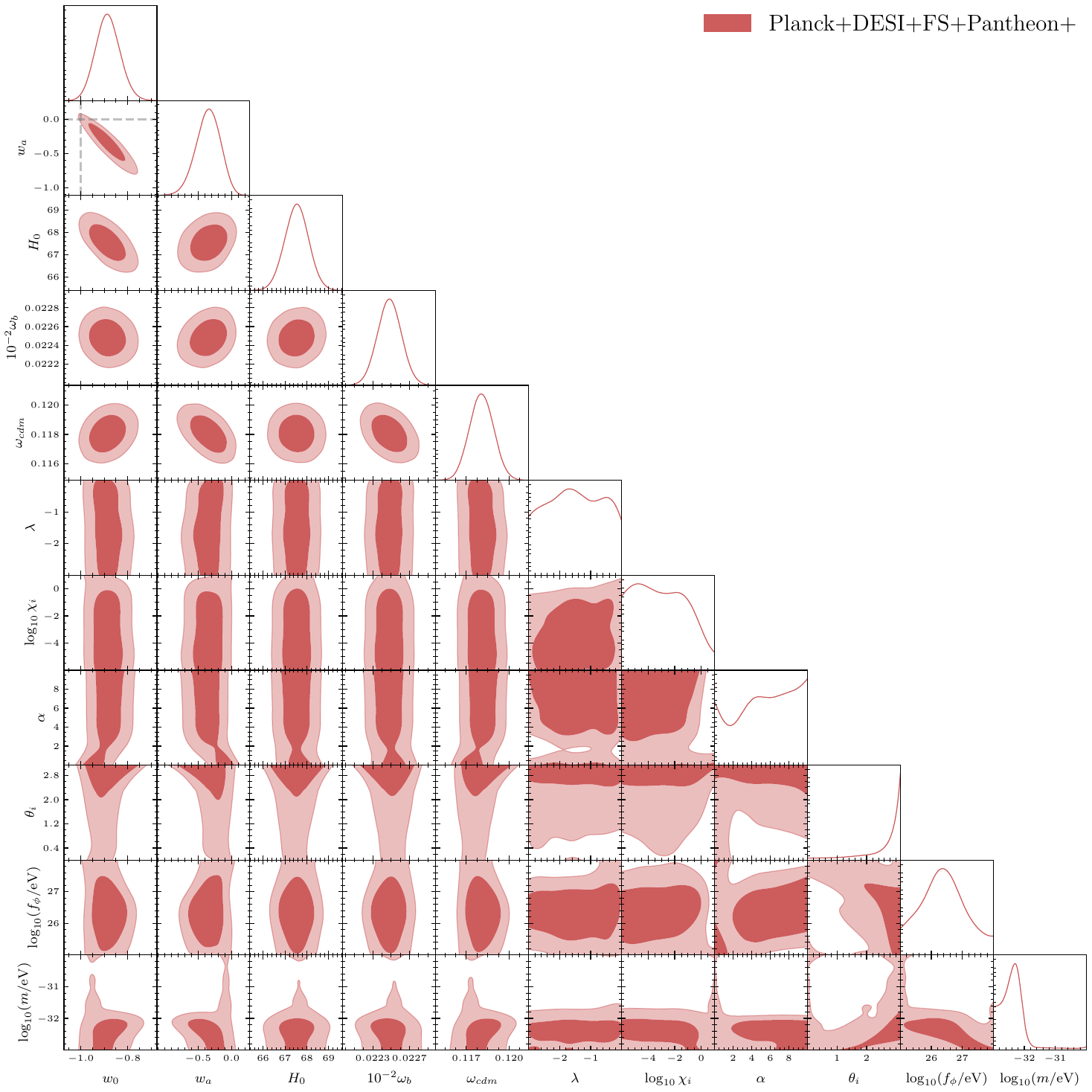}
    \caption{Full constraints at the 68\% and 95\% confidence levels for the kinetically mixed axion--dilaton (KMIX) model from the analysis~\cite{Chudaykin:2025aux,Chudaykin:2025auz}
    based on DESI DR2 BAO, \textit{Planck} CMB, Pantheon+ SNe, and DESI DR1 full-shape data.}
    \label{fig:placeholder_3}
\end{figure}

\begin{table*}[t]
\centering
\caption{
Summary of constraints for the CPL model (uniform prior in $w_0,w_a$) and the KMIX model.
For KMIX, $(w_0^{\rm (eff)},w_a^{\rm (eff)})$ denote the effective CPL parameters obtained by matching the KMIX expansion history to $H_{\rm CPL}(z)$ over the redshift range relevant for the data.
Quoted values are posterior means with 68\% minimum credible intervals, or one-sided 68\% limits where indicated.
}
\label{tab:cpl_kmix_compact}
\begin{tabular}{lccccc}
\hline\hline
 & CMB+DESI BAO & +Pantheon+ & +Union3 & +DES~Y5 & CMB+DESI FS \\
\hline
\multicolumn{6}{c}{\textbf{$w_0w_a$CDM}}\\
\hline
$\log(10^{10} A_\mathrm{s})$ 
& $3.033 \pm 0.014$
& $3.038 \pm 0.014$
& $3.035 \pm 0.014$
& $3.036 \pm 0.014$
& $3.035 \pm 0.013$ \\[3pt]

$n_\mathrm{s}$ 
& $0.9638 \pm 0.0038$
& $0.9650 \pm 0.0037$
& $0.9645 \pm 0.0037$
& $0.9646 \pm 0.0036$
& $0.9700 \pm 0.0036$ \\[3pt]

$\Omega_\mathrm{b} h^2$ 
& $0.02218 \pm 0.00013$
& $0.02223 \pm 0.00013$
& $0.02220 \pm 0.00013$
& $0.02221 \pm 0.00013$
& $0.02247 \pm 0.00013$ \\[3pt]

$\Omega_\mathrm{c} h^2$ 
& $0.11942 \pm 0.00087$
& $0.11887 \pm 0.00084$
& $0.11915 \pm 0.00084$
& $0.11906 \pm 0.00085$
& $0.11829 \pm 0.00083$ \\[3pt]

$w_0$ 
& $-0.45 \pm 0.21$
& $-0.843 \pm 0.055$
& $-0.681 \pm 0.089$
& $-0.757 \pm 0.057$
& $-0.868 \pm 0.052$ \\[3pt]

$w_a$ 
& $-1.64 \pm 0.59$
& $-0.59^{+0.21}_{-0.19}$
& $-1.03^{+0.30}_{-0.27}$
& $-0.83^{+0.23}_{-0.21}$
& $-0.42 \pm 0.18$ \\[3pt]

$H_0\,[\mathrm{km\,s^{-1}\,Mpc^{-1}}]$ 
& $63.8^{+1.8}_{-2.1}$
& $67.49 \pm 0.60$
& $65.96 \pm 0.85$
& $66.71 \pm 0.56$
& $67.53 \pm 0.57$ \\[3pt]

$\Omega_\mathrm{m}$ 
& $0.350 \pm 0.022$
& $0.3112 \pm 0.0058$
& $0.3266 \pm 0.0087$
& $0.3190 \pm 0.0056$
& $0.3102 \pm 0.0055$ \\[3pt]

$\sigma_8$ 
& $0.779^{+0.016}_{-0.018}$
& $0.8071 \pm 0.0087$
& $0.7958 \pm 0.0096$
& $0.8015 \pm 0.0084$
& $0.7999 \pm 0.0076$ \\[3pt]

\hline
\multicolumn{6}{c}{\textbf{KMIX}}\\
\hline
$\log(10^{10} A_\mathrm{s})$ 
& $3.034 \pm 0.014$
& $3.039 \pm 0.014$
& $3.036 \pm 0.014$
& $3.036 \pm 0.014$
& $3.037 \pm 0.013$ \\[3pt]

$n_\mathrm{s}$ 
& $0.9641 \pm 0.0038$
& $0.9655 \pm 0.0037$
& $0.9646 \pm 0.0037$
& $0.9648 \pm 0.0037$
& $0.9706 \pm 0.0036$ \\[3pt]

$\Omega_\mathrm{b} h^2$ 
& $0.02219 \pm 0.00013$
& $0.02224 \pm 0.00013$
& $0.02221 \pm 0.00013$
& $0.02222 \pm 0.00013$
& $0.02248 \pm 0.00013$ \\[3pt]

$\Omega_\mathrm{c} h^2$ 
& $0.11933 \pm 0.00088$
& $0.11869 \pm 0.00088$
& $0.11909 \pm 0.00086$
& $0.11897 \pm 0.00088$
& $0.11804 \pm 0.00084$ \\[3pt]

$w_0^{\rm (eff)}$ 
& $-0.50 \pm 0.21$
& $-0.855 \pm 0.059$
& $-0.691 \pm 0.089$
& $-0.762 \pm 0.059$
& $-0.885^{+0.049}_{-0.057}$ \\[3pt]

$w_a^{\rm (eff)}$ 
& $-1.51 \pm 0.60$
& $-0.52 \pm 0.22$
& $-0.99 \pm 0.30$
& $-0.80 \pm 0.24$
& $-0.35^{+0.21}_{-0.16}$ \\[3pt]

$H_0\,[\mathrm{km\,s^{-1}\,Mpc^{-1}}]$ 
& $64.2^{+1.8}_{-2.1}$
& $67.47 \pm 0.60$
& $65.99 \pm 0.83$
& $66.68 \pm 0.56$
& $67.54 \pm 0.56$ \\[3pt]

$\Omega_\mathrm{m}$ 
& $0.345 \pm 0.021$
& $0.3111 \pm 0.0058$
& $0.3261 \pm 0.0086$
& $0.3190 \pm 0.0056$
& $0.3096 \pm 0.0055$ \\[3pt]

$\sigma_8$ 
& $0.782 \pm 0.017$
& $0.8058 \pm 0.0088$
& $0.7957 \pm 0.0094$
& $0.8008 \pm 0.0085$
& $0.7987 \pm 0.0075$ \\[3pt]

$\lambda$ 
& $-1.3^{+1.3}_{-1.5}$
& $-1.45^{+1.2}_{-0.72}$
& $-1.45^{+1.2}_{-0.72}$
& $-1.46^{+1.1}_{-0.72}$
& $> -1.76$ \\[3pt]

$\log_{10}\chi_i$ 
& $-2.9^{+3.8}_{-3.1}$
& $< -1.75$
& $-3.0^{+3.9}_{-3.0}$
& $< -1.95 $
& $< -1.75$ \\[3pt]

$\alpha$ 
& $> 5.28$
& $> 4.71$
& $> 5.37$
& $> 5.27$
& $> 3.78$ \\[3pt]

$\theta_i$ 
& $> 3.11$
& $> 3.07$
& $3.085^{+0.041}_{+0.0033}$
& $> 3.11$
& $> 2.76$ \\[3pt]

$\log_{10}(f_\phi/\mathrm{eV})$ 
& $25.65^{+0.23}_{-0.48}$
& $25.85^{+0.30}_{-0.49}$
& $25.68^{+0.26}_{-0.47}$
& $25.72^{+0.27}_{-0.45}$
& $26.06^{+0.42}_{-0.62}$ \\[3pt]

$\log_{10}(m/\mathrm{eV})$ 
& $-32.16^{+0.17}_{-0.054}$
& $-32.24^{+0.24}_{-0.15}$
& $-32.18^{+0.18}_{-0.053}$
& $-32.20^{+0.19}_{-0.061}$
& $-32.29^{+0.25}_{-0.40}$ \\[2pt]
\hline\hline
\end{tabular}
\end{table*}

\end{document}